\documentclass[aps,prd,twocolumn,superscriptaddress,amsmath,widetext,amssymbepsfig,showpacs,nofootinbib]{revtex4}
\usepackage{graphicx}
\usepackage{dcolumn}
\usepackage{bm}
\usepackage{natbib}

\newcommand{\be}{\begin{equation}}
\newcommand{\ee}{\end{equation}}
\newcommand{\bea}{\begin{eqnarray}}
\newcommand{\eea}{\end{eqnarray}}

\newcommand{\Mpc}{{\rm ~Mpc}}

\urldef\mgcamb\url{http://www.sfu.ca/~aha25/MGCAMB.html}

\begin{document}
\title{New constraints on Coupled Dark Energy from Planck}

\author {Valentina Salvatelli}
\affiliation{Physics Department and INFN, Universit\`a di Roma ``La Sapienza'', Ple Aldo Moro 2, 00185, Rome, Italy}
\author {Andrea Marchini}
\affiliation{Physics Department and INFN, Universit\`a di Roma ``La Sapienza'', Ple Aldo Moro 2, 00185, Rome, Italy}
\author {Laura Lopez-Honorez}
\affiliation{Theoretische Natuurkunde, Vrije Universiteit Brussel and The International Solvay Institutes Pleinlaan 2, B-1050 Brussels, Belgium}
\author {Olga Mena}
\affiliation{IFIC, Universidad de Valencia-CSIC, 46071, Valencia,  Spain}

\begin {abstract}
We present new constraints on coupled dark energy from the recent measurements of the Cosmic Microwave Background
Anisotropies from the Planck satellite mission. We found that a coupled dark energy model is fully compatible with 
the Planck measurements, deriving a weak bound on the dark matter-dark energy coupling parameter $\xi=-0.49^{+0.19}_{-0.31}$ at $68 \%$ c.l.. Moreover if Planck data are fitted to a coupled dark energy scenario, the constraint on the Hubble constant is relaxed to
$H_0=72.1^{+3.2}_{-2.3}$ km/s/Mpc, solving the tension with the Hubble Space Telescope value.
We show that a combined Planck+HST analysis provides significant evidence for coupled dark energy finding a non-zero value 
for the coupling parameter $\xi$, with $-0.90< \xi <-0.22$ at $95\%$ c.l.. We also consider the combined constraints from the Planck data plus the BAO measurements of the 6dF Galaxy Survey, the Sloan Digital Sky Survey and the Baron Oscillation Spectroscopic Survey.
\end {abstract}


\maketitle

\section {Introduction} \label{sec:intro}

The Planck satellite experiment has recently provided new and precise
measurements of the Cosmic Microwave Background (CMB) anisotropy,
covering a wide range of angular scales up to multipole $\ell \sim
2500$ \cite{Ade:2013xsa, Ade:2013lta, Planck:2013kta}.  The new data 
are in full agreement with the expectations of the so-called
standard $\Lambda$CDM scenario.

There exist however some tensions when the value of a number of cosmological
parameters, as measured by Planck data, are compared 
with the values of  the same parameters as measured by independent
cosmological probes.

The most notable case concerns the constraint on the Hubble constant
$H_0$. The value measured by the Planck team
, $H_0=67.3\pm1.2$ km/s/Mpc (at $68 \%$ c.l.), is significantly lower than the previous measurement 
of $H_0=73.8\pm2.4$ km/s/Mpc (at $68 \%$ c.l.) from the Hubble Space
Telescope (HST) arising from optical and 
infrared observations of $\sim 600$ Cepheid variables
~\cite{riess2011}. While systematics can certainly be present in both measurements, it is timely to investigate
if this discrepancy can be explained by the inclusion of new physical phenomena.

Very recently, Marra et al.~\cite{marra}, have pointed out that the HST value could be affected by
the cosmic variance of the local expansion rate and inhomogeneities. In this approach,
the HST value is therefore biased and the correct value is given by
Planck measurements.

On the other hand, one has to consider that the CMB determination of the Hubble parameter is
not a direct measurement but it is based on the assumption of an underlying theoretical model.

For example, the inclusion in the Planck data analysis of an extra
relativistic energy component, parameterized via the effective number
of relativistic degrees of freedom $N_\textrm{eff}$, changes the
constraint on the Hubble constant to $H_0=70.7\pm3.2$ km/s/Mpc at $68
\%$ c.l. \cite{Ade:2013lta}, totally compatible with the HST
determination.  Several physical candidates have been recently
examined to account for the extra dark radiation component (see
e.g. \cite{eleonora} and references therein). A combined Planck and
HST data analysis yields  evidence for dark radiation at more than
$95 \%$ c.l. (\cite{Ade:2013lta},\cite{eleonora2}).

In this paper we investigate the possibility that the solution
of the Planck-HST tension could arise from an interaction between dark energy and
dark matter. The assumption of a coupled dark energy model can change
significantly the CMB constraints on the Hubble parameter (see e.g. 
\cite{Olivares:2005tb,Valiviita:2009nu,Martinelli:2010rt,DeBernardis:2011iw}). In particular,
if coupled models are fitted to a (wrong) $\Lambda$CDM cosmology, the reconstructed 
$H_0$ from CMB will be shifted from its real value~\cite{Honorez:2010rr}.
A mismatch between low and high redshift $H_0$ measurements could
therefore be a smoking gun for an interaction in the dark sector.

Moreover, even if a cosmological constant $\Lambda$ is a good fit to the current data,
the dark matter-dark energy interaction alleviates the well-known coincidence problem 
that plagues the $\Lambda$CDM scenario (see e.g. \cite{Peebles:2002gy}). Several candidates 
for coupled dark energy models have been proposed and investigated (see e.g. \cite{models}).
Since coupled quintessence models can resemble to scalar-tensor or brans dicke gravitational
theories, it has been suggested that coupled dark energy can also hint to a modification of
general relativity on cosmological scales (see e.g. \cite{Amendola:1999er,Farrar:2003uw,Pettorino:2008ez}).

The paper is organized as follows. In Section~\ref {sec:theories} we briefly present the coupled dark energy model we consider for
 our analysis, in Sec.~\ref{sec:analysis} we describe the analysis method, in Sec.~\ref {sec:constraints} we show our results and 
in Section\ref {sec:concl} we draw our conclusions.

\section{Dark interaction parametrization} \label{sec:theories}

We assume a flat universe described by the Friedmann-Robertson-Walker
metric. We parametrize the interaction as follows:
\begin{eqnarray}
  \label{eq:conservDM}
\nabla_\mu T^\mu_{(dm)\nu} &=&Q \,u_{\nu}^{(dm)}/a~, \\
  \label{eq:conservDE}
\nabla_\mu T^\mu_{(de)\nu} &=&-Q \,u_{\nu}^{(dm)}/a~, 
\end{eqnarray}
where $T^\mu_{(dm)\nu}$ and $T^\mu_{(de)\nu}$ are the energy-momentum tensors for the dark matter and dark energy components respectively, $u_{\nu}^{(dm)}$ is the dark matter four-velocity and the coefficient $Q$ encodes the interaction rate between the two dark components. 

In particular we restrict the parametrization to the case where the
interaction rate is proportional to the dark energy density $\rho_{de}$: 
\begin{equation}
Q=\xi \mathcal{H} \rho_{de}
\label{rate}
\end{equation}
where $\xi$ is a dimensionless parameter and $\mathcal{H}=\dot{a}/a$ (the dot
indicates derivative respect to conformal time
$d\tau=dt/a$).  Such an interacting model is in agreement
with cosmological constraints and it does not suffer from early time
instabilities if the coupling $\xi$ is negative and the dark
energy equation of state $w$ satisfies $w>-1$~\cite{He:2008si,Gavela:2009cy}. 
We shall follow here the former stability conditions.

The background evolution equations in the coupled model considered
here read~\cite{Gavela:2010tm}
\begin{eqnarray}
  \label{eq:backDM}
\dot{{\rho}}_{dm}+3{\mathcal H}{\rho}_{dm}= \xi{\mathcal H}{\rho}_{de}~, \\
  \label{eq:backDE}
\dot{{\rho}}_{de}+3{\mathcal H}(1+w){\rho}_{de}= -\xi{\mathcal
  H}{\rho}_{de}~.
\end{eqnarray}  
In the synchronous gauge, the evolution of the dark matter and dark energy perturbations in the linear regime reads~\cite{Gavela:2010tm}
 \begin{eqnarray}
\label{eq:deltambe}
\dot\delta_{dm}  & = & -(k v_{dm}+\frac12 \dot h) +\xi {\mathcal H}\frac{\rho_{de}}{\rho_{dm}} \left(\delta_{de}-\delta_{dm}\right)
\\ \nonumber
&& +\xi \frac{\rho_{de}}{\rho_{dm}} \left(\frac{k v_T}{3}+\frac{\dot h}{6}\right)\,, \\
\label{eq:deltaees}
\dot\delta_{de}  & = & -(1+w)(k v_{de}+\frac12 \dot h)-3 {\mathcal H}\left(1 -w\right)
   \\ \nonumber
&& \left[ \delta_{de} +{\mathcal H} \left( 3(1+w) + \xi\right)\frac{v_{de}}{k} \right]-\xi \left(\frac{k v_T}{3}+\frac{\dot h}{6}\right) \,,\\
\label{eq:thetames}
 \dot v_{dm}  & = & -{\mathcal H} v_{dm} \,,\\
 \label{eq:thetaees}
\dot v_{de}  & = & 2 {\mathcal H}\left(1 +\frac{\xi}{1+w} \right)
    v_{de}+\frac{k}{1+w}\delta_{de}-\xi{\cal H}\frac{v_{dm}}{1+w}\,,
\end{eqnarray}
where $\delta_{dm,de}$ and $v_{dm,de}$ are the density perturbations
and velocities of the dark matter and dark energy fluids,
respectively, $v_T$ is the center of mass velocity for the total
fluid and $h$ is the usual synchronous gauge metric
perturbation.  Equations~(\ref{eq:deltambe})-(\ref{eq:thetaees})
include the contributions of the perturbation in the expansion rate
$H={\mathcal H}/a + \delta H$, the dark energy
speed of sound has been fixed to 1, i.e. $\hat c_{s\,de}^2=1$, and the
equation of state for dark energy $w$ has been taken to be
constant. 

Also notice that in our numerical analysis, we have
considered adiabatic initial conditions for all components, see
appendix \ref{appendix}.

\begin{figure}[htb!]
\centering
\includegraphics[width=9cm]{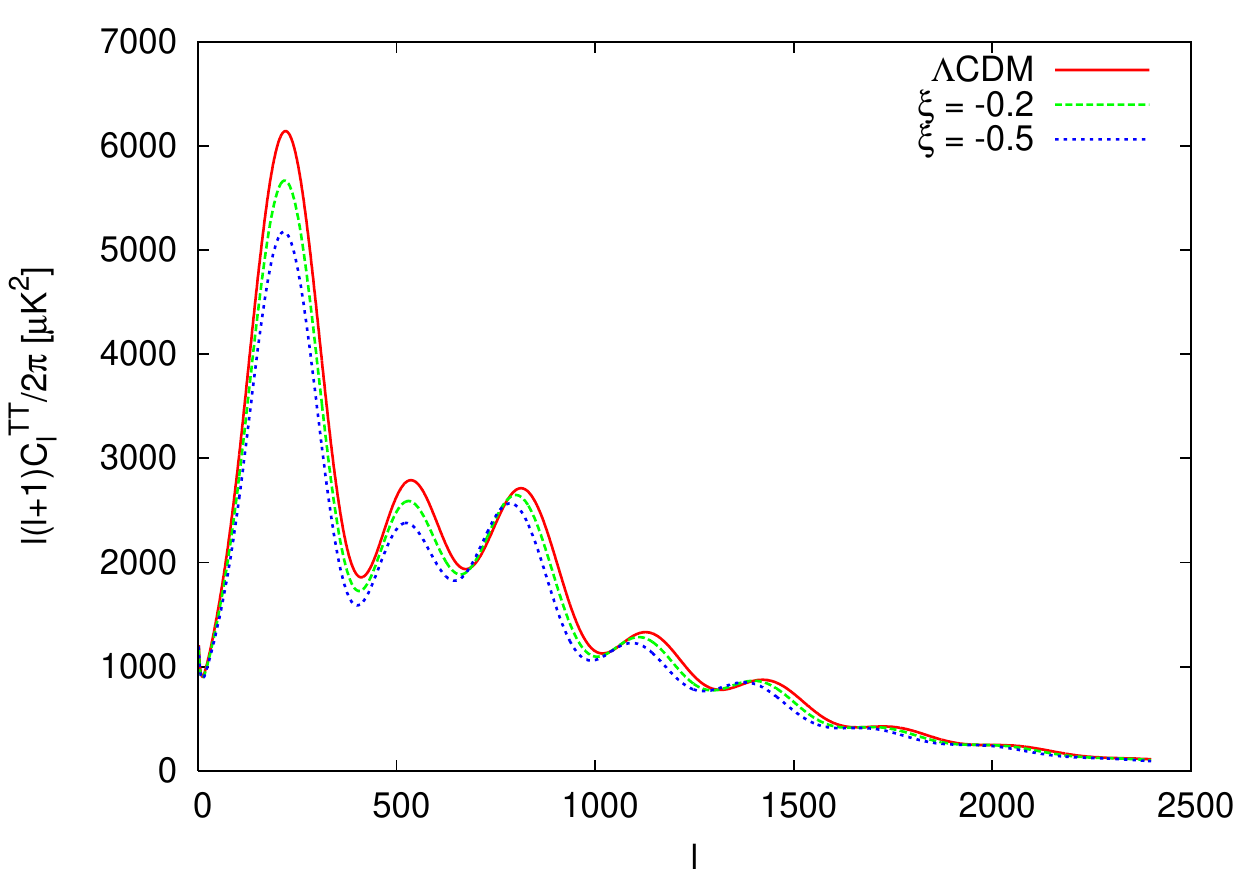}
\caption{CMB temperature power spectrum in the $\Lambda$CDM case and in the coupled cases for $\xi=-0.2, -0.5$, $\Omega_c h^2= 0.1186$, $H_0=67.9$ km/s/Mpc. The main effects of the coupling are shifting the position of the acoustic peaks and varying their amplitude.}
\label{figSpectra}
\end{figure}

The dark interaction we consider affects the CMB temperature spectrum
in several ways. In Fig.~\ref{figSpectra}, we illustrate the impact of 
$\xi$ up to multipole $l=2500$ for $\xi=-0.2, -0.5$ assuming a cold 
dark matter density $\Omega_c h^2= 0.1186$ and $H_0=67.9$ km/s/Mpc. Notice that the presence of
a coupling among the dark matter and the dark energy fluids shifts the 
position of the peaks to larger multipoles. At low multipoles, a value of
$\xi$ different from zero contributes to the late integrated Sachs-Wolfe 
(ISW) effect, while at high multipoles changes the amplitude of the gravitational lensing.

\section {Data Analysis Method} \label{sec:analysis}

The theoretical CMB angular spectra are computed with a modified version of the CAMB code \cite{Lewis:1999bs}, in which we have included
the background and linear density perturbation equations in the presence of a coupling between the dark matter and the dark energy sectors.
\begin{table*}[htb!]
\begin{center}
\begin{tabular}{|l|c|}
\hline
\hline
Parameters & Prior  \\ \hline
$\Omega_b h^2$ & [0.005, 0.100] \\
$\Omega_c h^2$ & [0.005, 0.100] \\
$100\theta$ & [0.5, 10] \\
$\tau$ & [0.01, 0.80] \\
$n_s$ &  [0.9, 1.1]\\
$\log(10^{10} A_s)$ & [2.7, 4.0] \\
$\xi$ & [-1,0]\\
\hline
\hline
\end{tabular}
\caption{Ranges for the priors of different cosmological parameters considered in the analysis.}
\label{Tab2}
\end{center}
\end{table*}
For the Planck data set\footnote{Planck data set is publicly available at
  \textit{http://pla.esac.esa.int/pla/aio/planckProducts.html}}, we consider the high-$\ell$ TT likelihood, including measurements up to a 
maximum multipole number of $\ell_{\rm max}=2500$, combined with the
low-$\ell$ TT likelihood which includes measurements up to $\ell=49$. We also consider
low-$\ell$ ($\ell<23$) TE,EE,BB likelihood which includes polarization measurements from nine years of observation of the Wilkinson
Microwave Anisotropy Probe (WMAP)~\cite{Bennett:2012fp}. We refer to this data combination as the PLANCK data set.

We also consider the effect of a gaussian prior on the Hubble constant $H_0=73.8\pm2.4 \,\mathrm{km}\,\mathrm{s}^{-1}\,\mathrm{Mpc}^{-1}$
consistent with the measurements of the Hubble Space Telescope as in \cite{riess2011}. We refer to this prior as HST.

Finally we analyze the impact of baryon acoustic oscillations (BAO) measurements, using the data of three surveys
at different redshifts: the 6dF Galaxy Survey measurement at $z = 0.1$ provided in~\cite{Beutler:2011hx},
the SDSS DR7 measurement at $z=0.35$ from~\cite{Padmanabhan:2012hf} and the
BOSS DR9 measurement at $z = 0.57$ discussed in~\cite{Anderson:2012sa}. We refer to this
combination as the BAO data set.

We sample a seven-dimensional set of cosmological parameters, adopting
flat priors on them (see Tab.~\ref{Tab2}): the $\xi$ coupling parameter, the baryon 
and cold dark matter densities $\Omega_{\rm b} h^2$ and $\Omega_{\rm c} h^2$, the ratio of the sound horizon 
to the angular diameter distance at decoupling $\theta$, the optical
depth to reionization $\tau$, the scalar spectral index $n_s$ and the
amplitude of the primordial scalar perturbation spectrum, $A_s$ at $k=0.05\Mpc^{-1}$. 
We fix the relativistic number of degrees of freedom parameter to
$N_{eff}=3.046$, the helium abundance to $Y_p=0.24$, the total
neutrino mass to $\sum m_{\nu}=0.06 eV$, the spectrum lensing
normalization to $A_L=1$ and the dark energy equation of state to $w=-0.999$.
We marginalize over all foregrounds parameters as described in~\cite{Ade:2013lta}.
We also consider the effect of letting $w$ free to vary under the condition $w>-1$.

Our analysis follows a Markov chains approach and is based on a
modified version of the public available code CosmoMC~\cite{Lewis:2002ah, Lewis:2013hha}, setting the statistical convergence
for Gelman and Rubin $R-1$ values below $0.03$.

\section{Results} \label {sec:constraints}

We have initially performed four runs, fixing the dark energy equation of state to $w=-0.999$: 
a first run with no coupling, $\xi=0$, which corresponds to the $\Lambda$CDM case (as in~\cite{Ade:2013lta}). 
We have then let the coupling parameter to vary freely, performing separately an analysis with
Planck data alone, Planck data plus the HST prior on the Hubble constant and Planck data plus BAO data. 

\begin{figure*}[htb!]
\centering
\includegraphics[width=19cm]{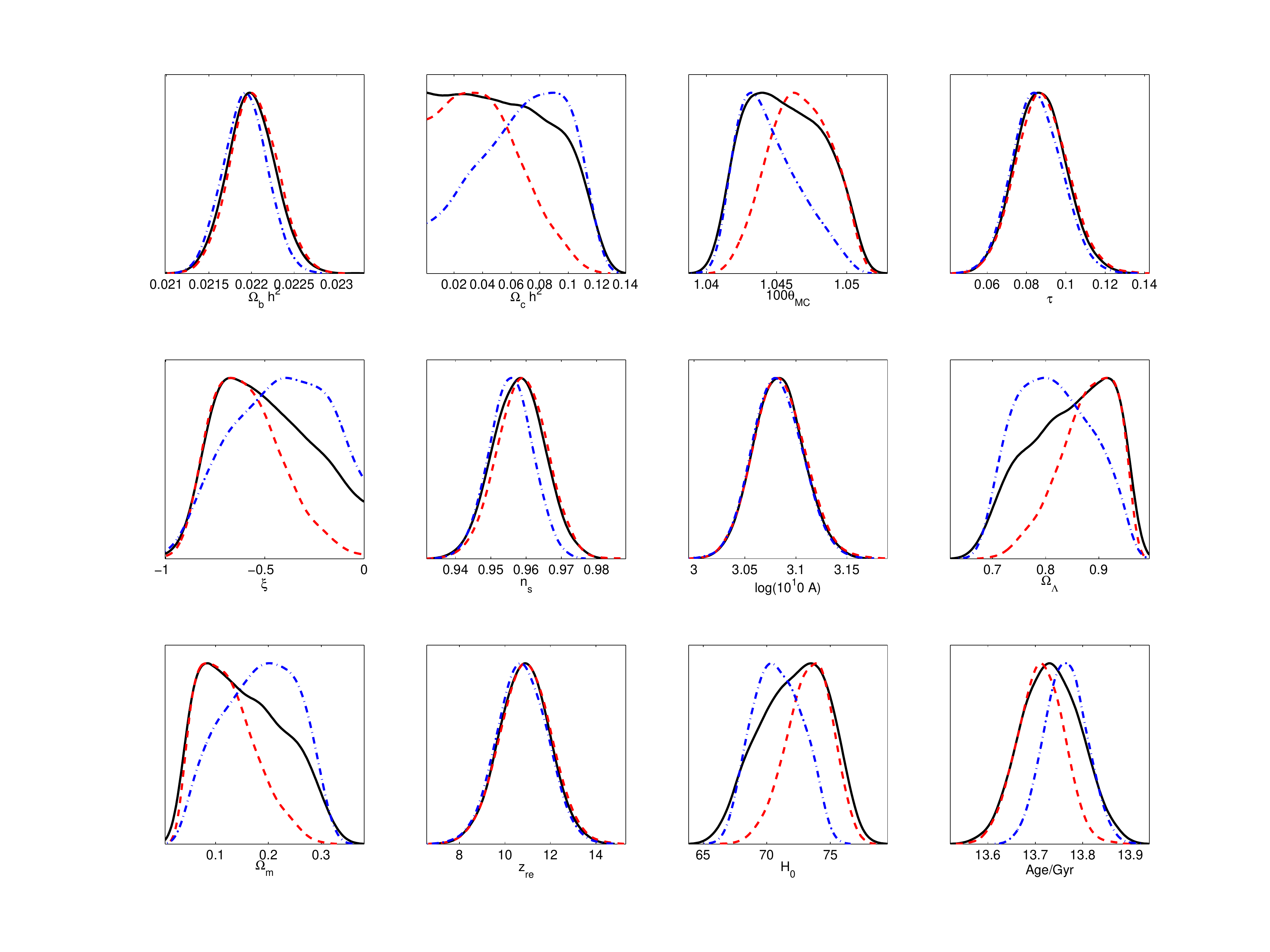}
\caption{Posterior distributions for the cosmological parameters presented in Tab.\ref{Tab1} from PLANCK data set alone (solid black line), PLANCK plus HST prior (red dashed line) and PLANCK plus BAO measurements (blue dot-dashed line). The effect of HST prior is to provide narrower distributions and stronger constraints, especially for the coupling parameter $\xi$.}
\label{fig.posterior_wfix}
\end{figure*}

\begin{figure*}[htb!]
\includegraphics[width=4.0cm]{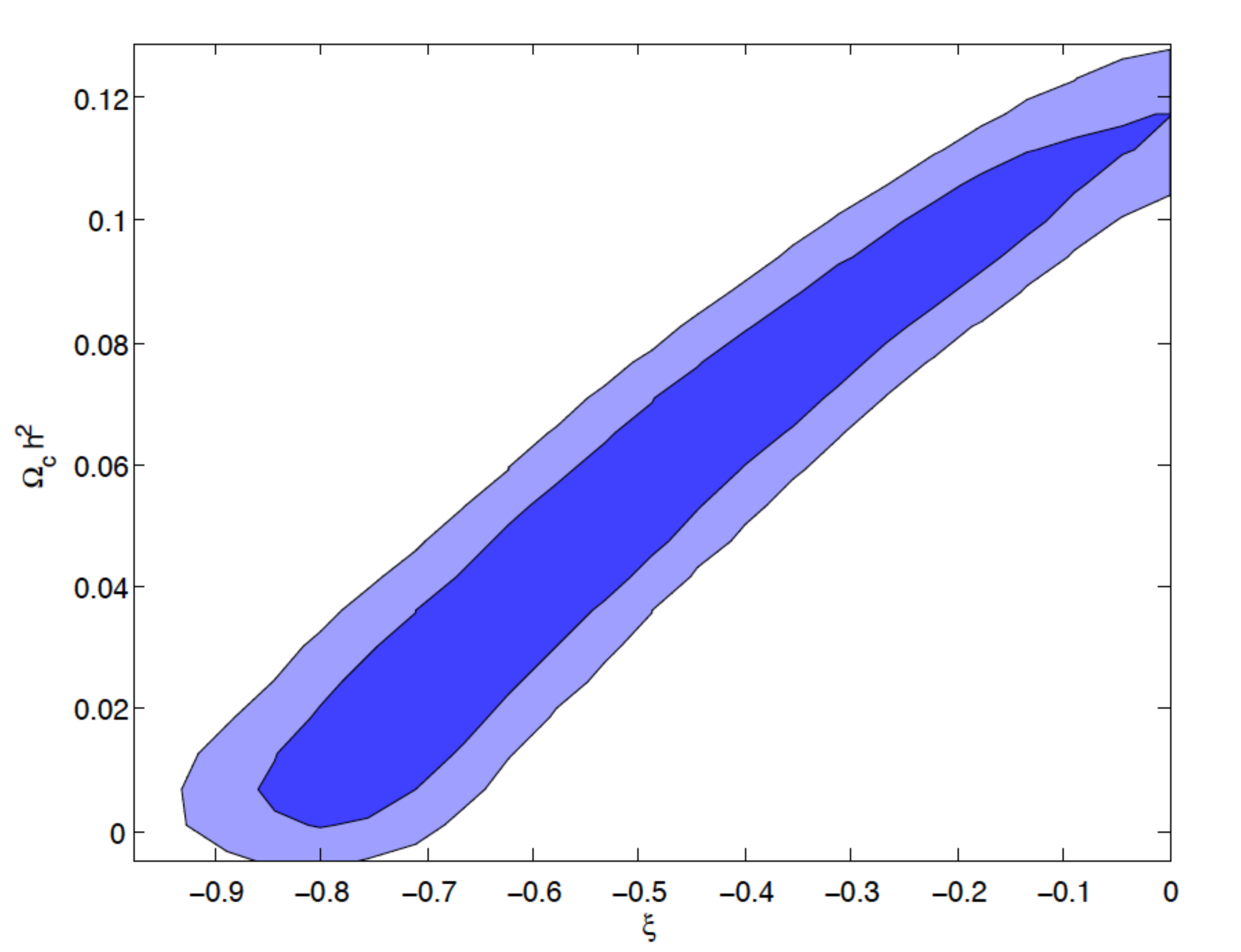}
\includegraphics[width=4.0cm]{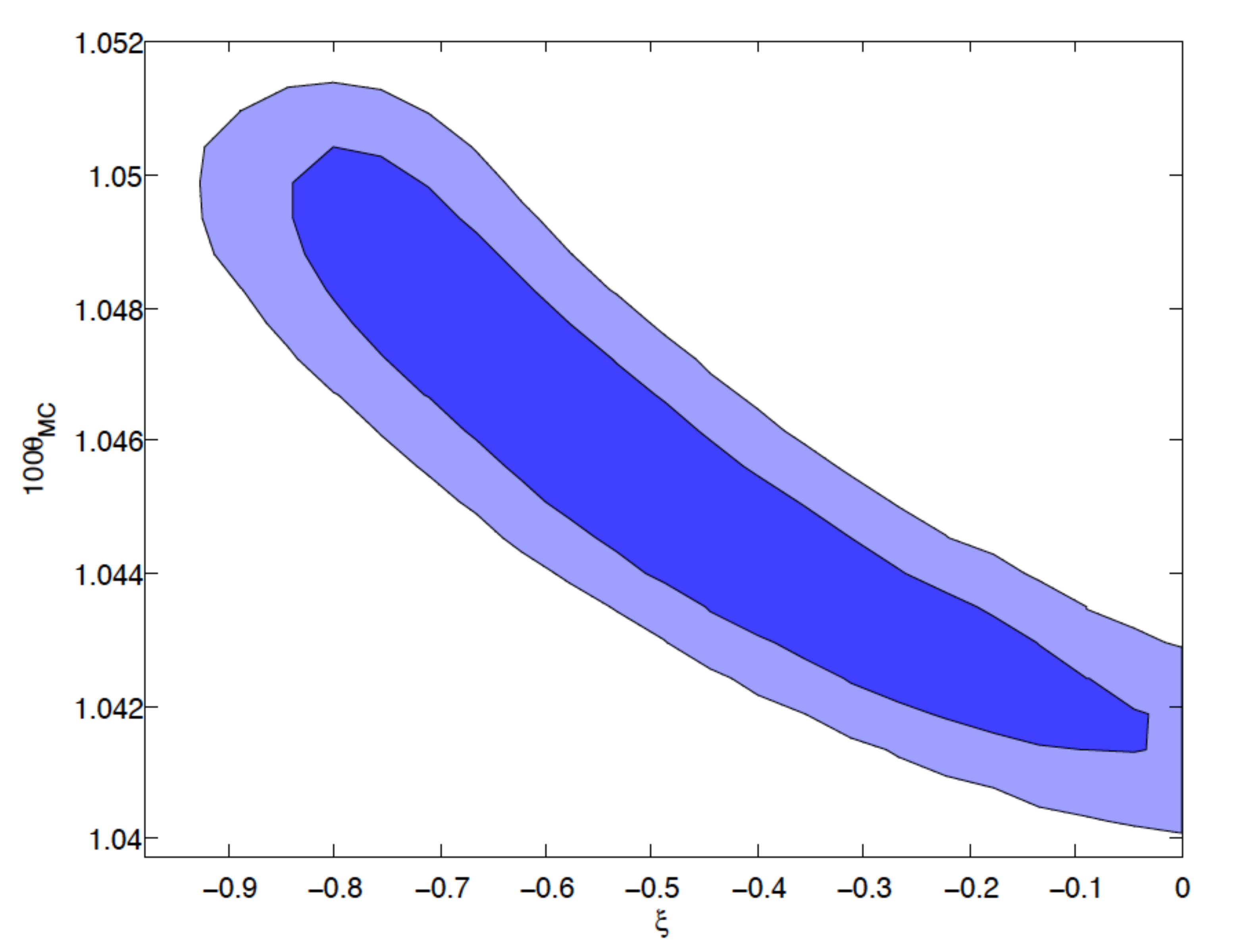}
\includegraphics[width=4.0cm]{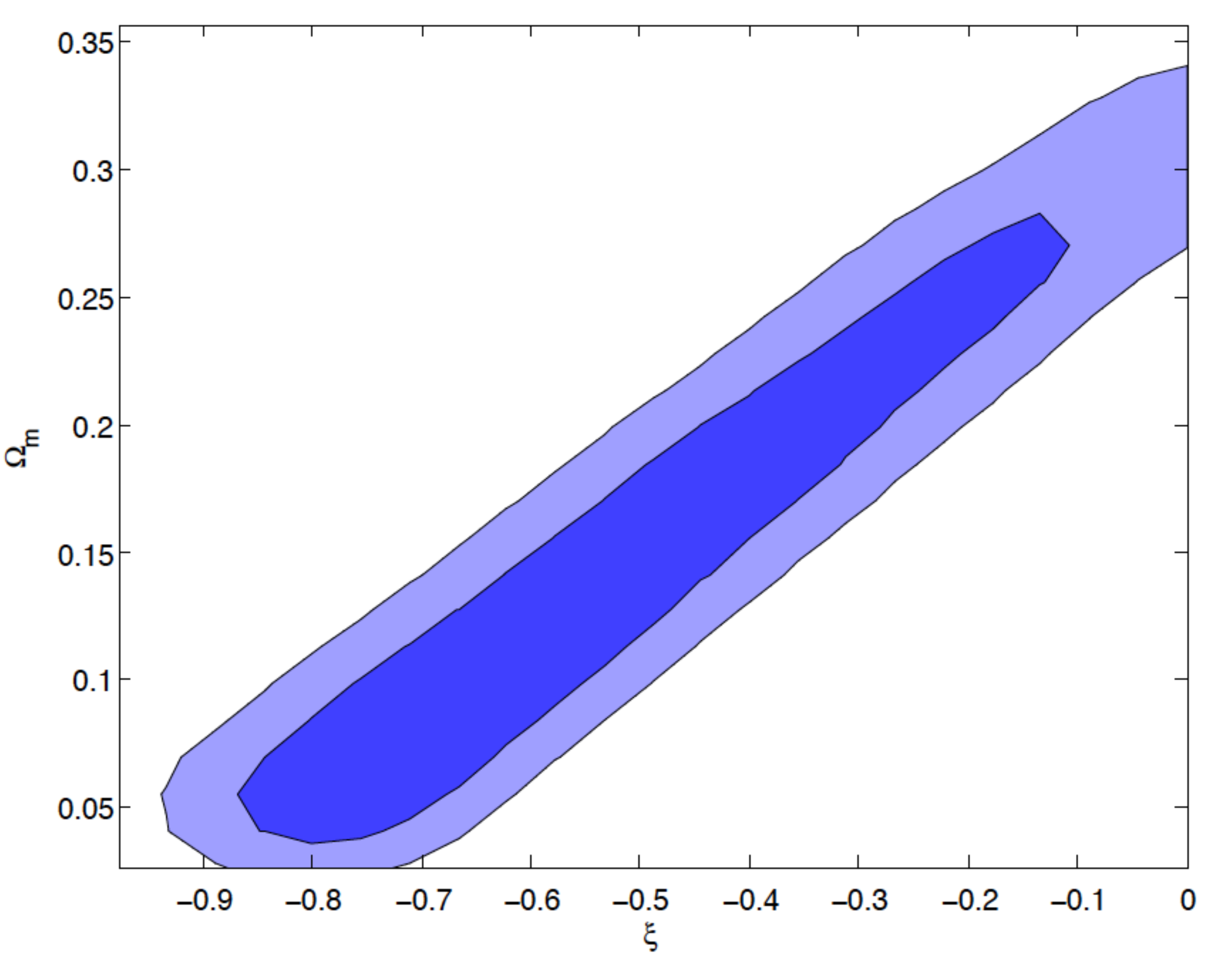}
\includegraphics[width=4.0cm]{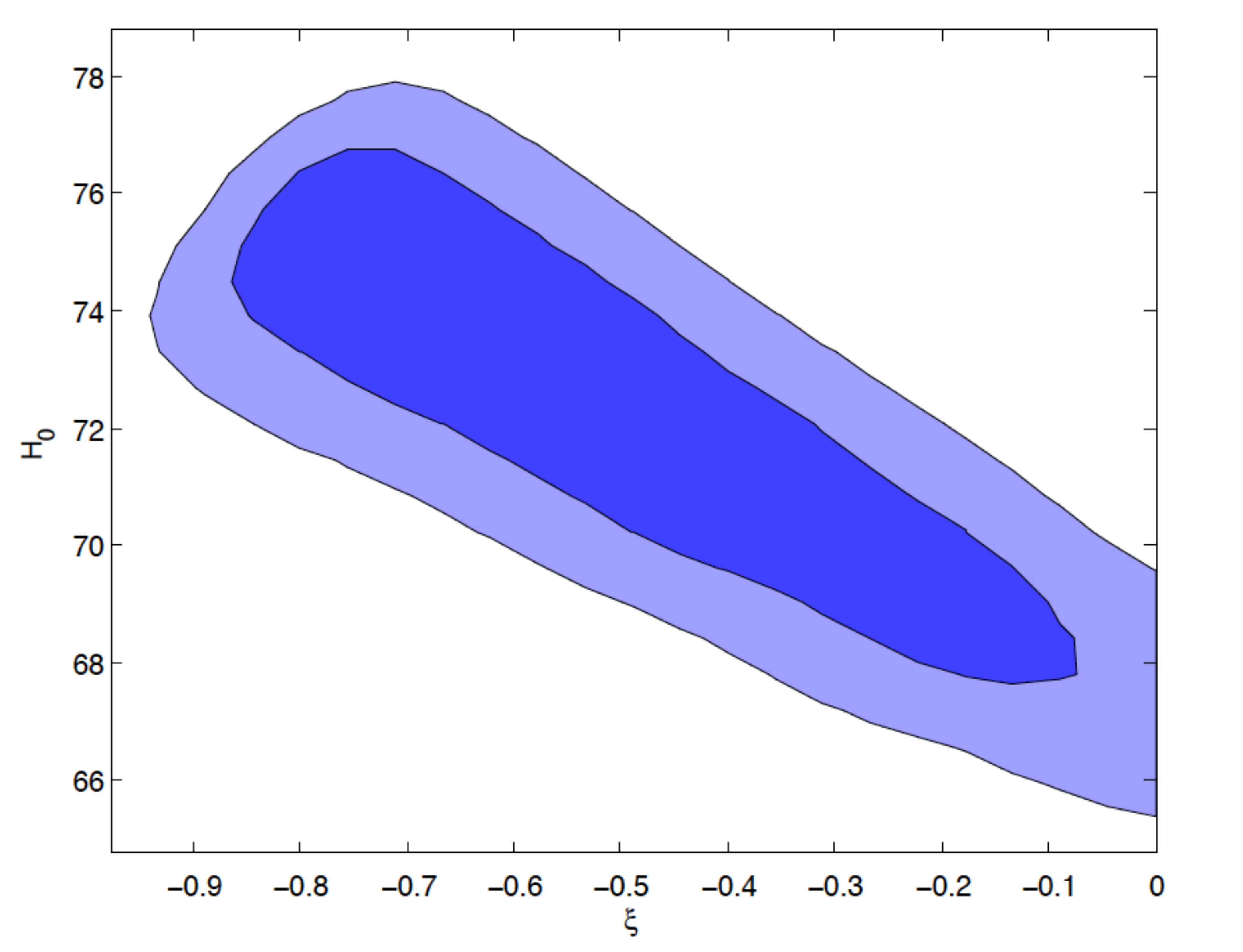}
\caption{2-D posterior distributions of parameters most degenerate with the coupling $\xi$. A strong correlation is evident with the cold dark matter density parameter. A larger absolute value of the coupling $\xi$ implies a decrement of the dark cold matter and a consequent decrease of the dark matter density. Since the assumption of a flat universe, it also implies a larger dark energy amount that brings to an increment of $H_0$ and consequently an increase of $\theta$.}
\label{fig.2Dcontour}
\end{figure*}

\begin{figure*}[htb!]
\includegraphics[width=4.0cm]{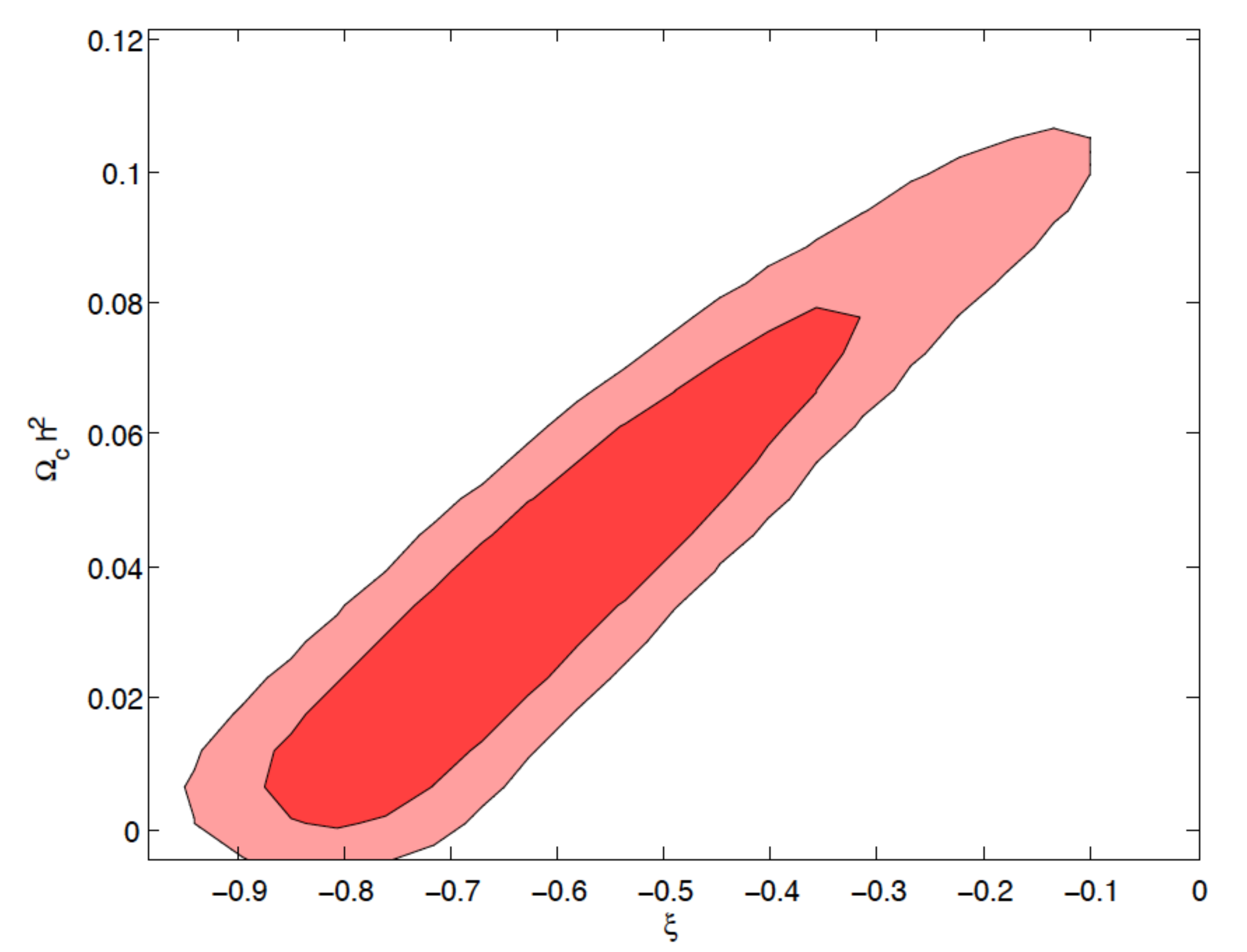}
\includegraphics[width=4.0cm]{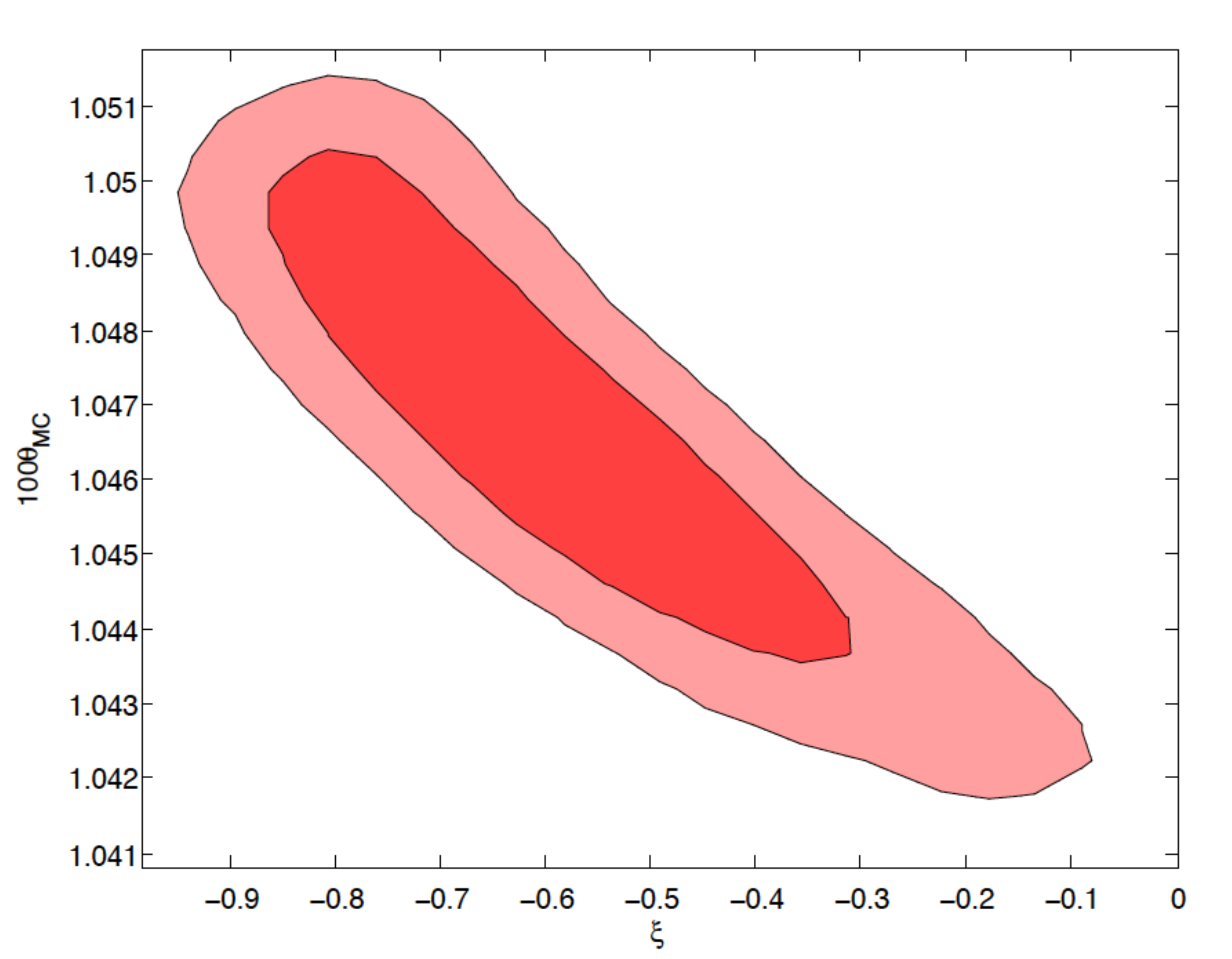}
\includegraphics[width=4.0cm]{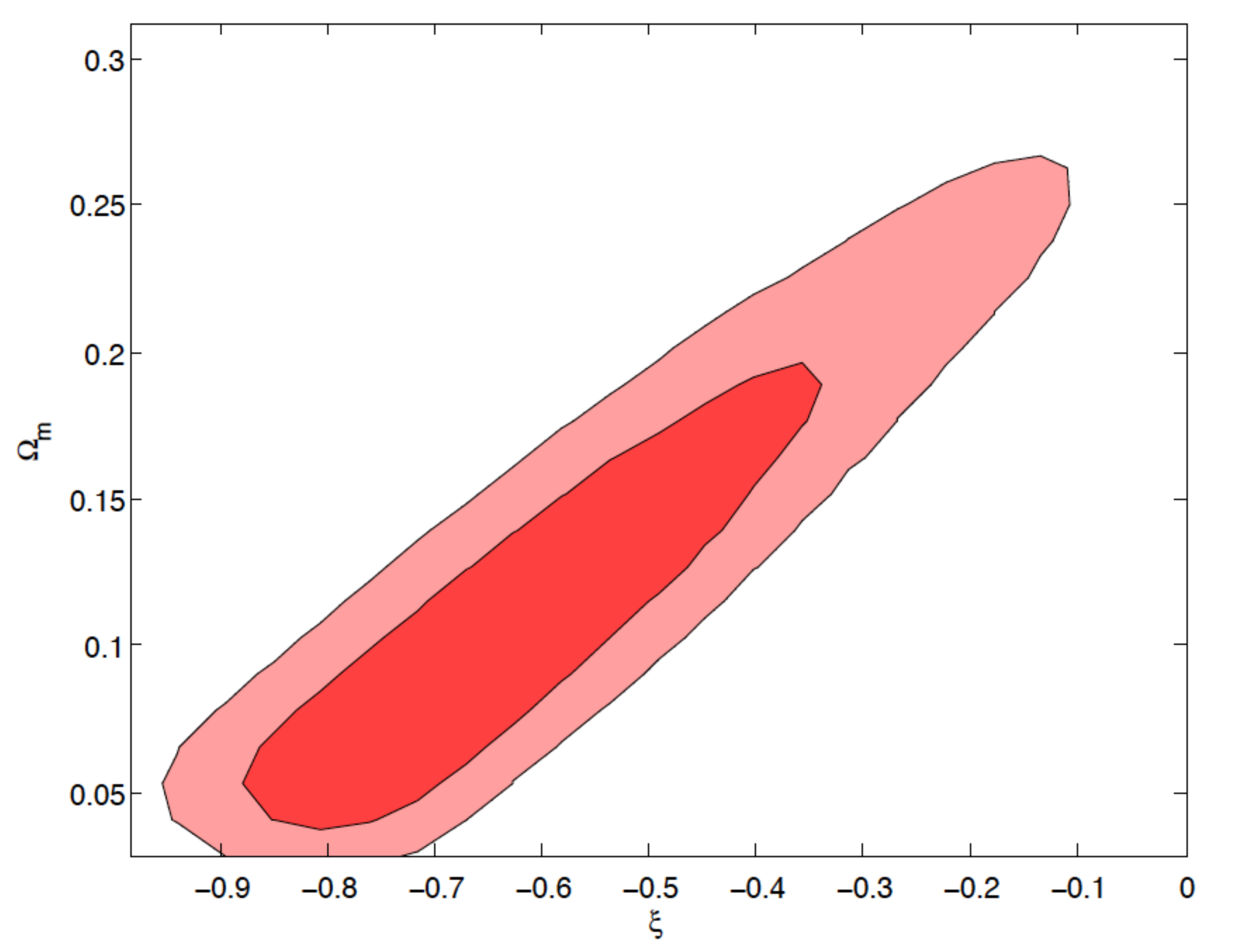}
\includegraphics[width=4.0cm]{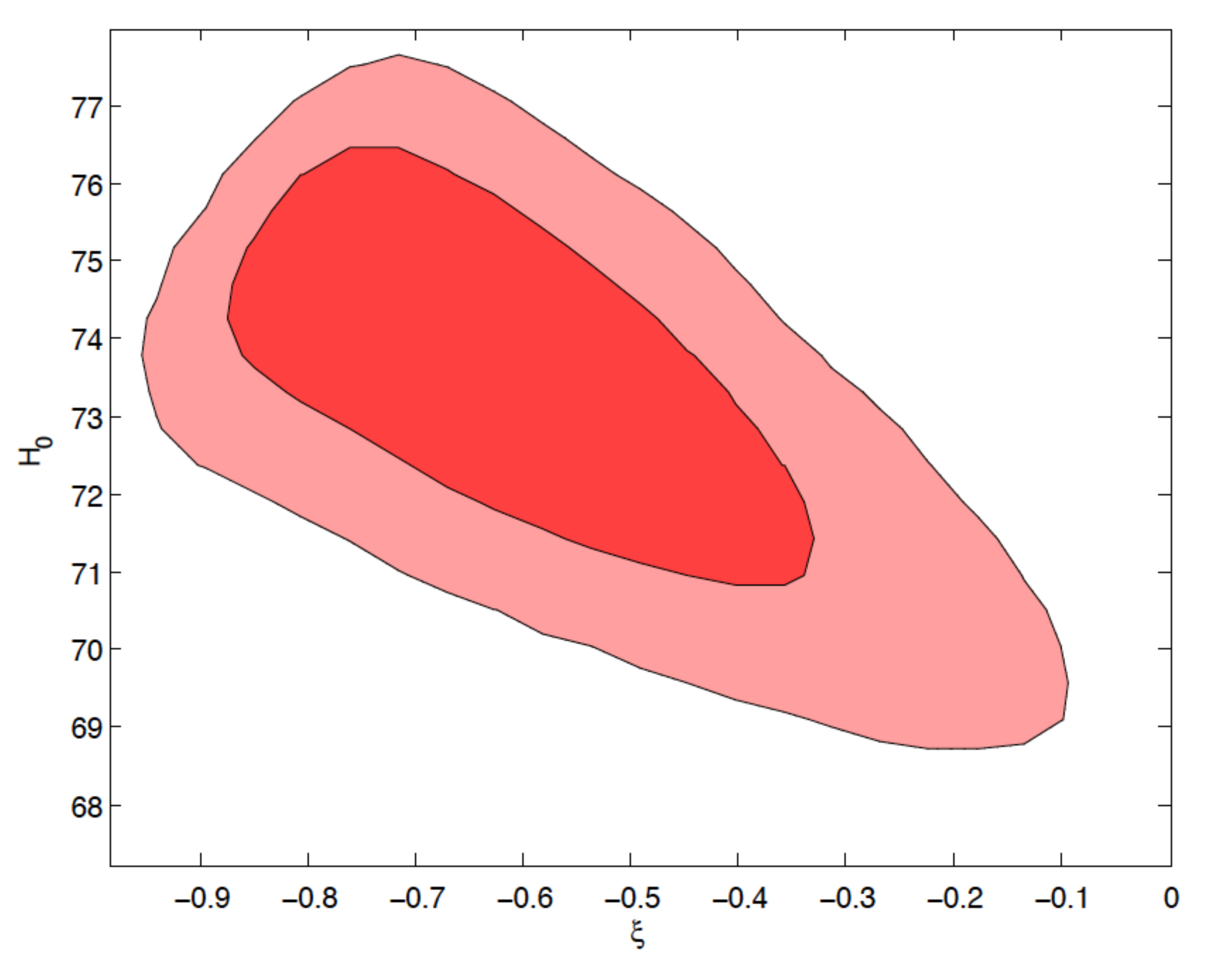}
\caption{2-D posterior distributions  in presence of the HST prior for the same parameters shown in Fig.\ref{fig.2Dcontour}.}
\label{fig.2DcontourHST}
\end{figure*}

\begin{figure*}[htb!]
\includegraphics[width=4.0cm]{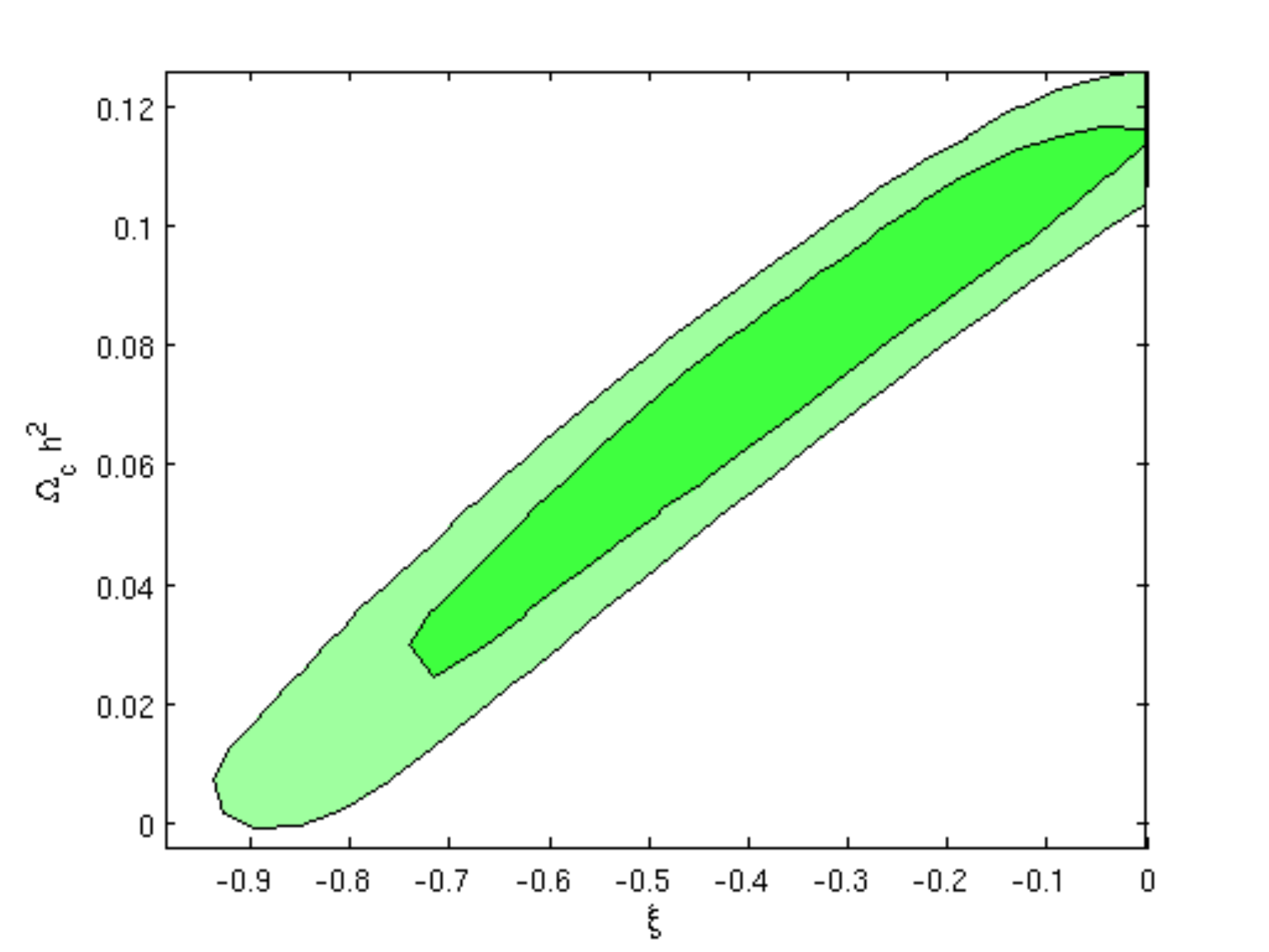}
\includegraphics[width=4.0cm]{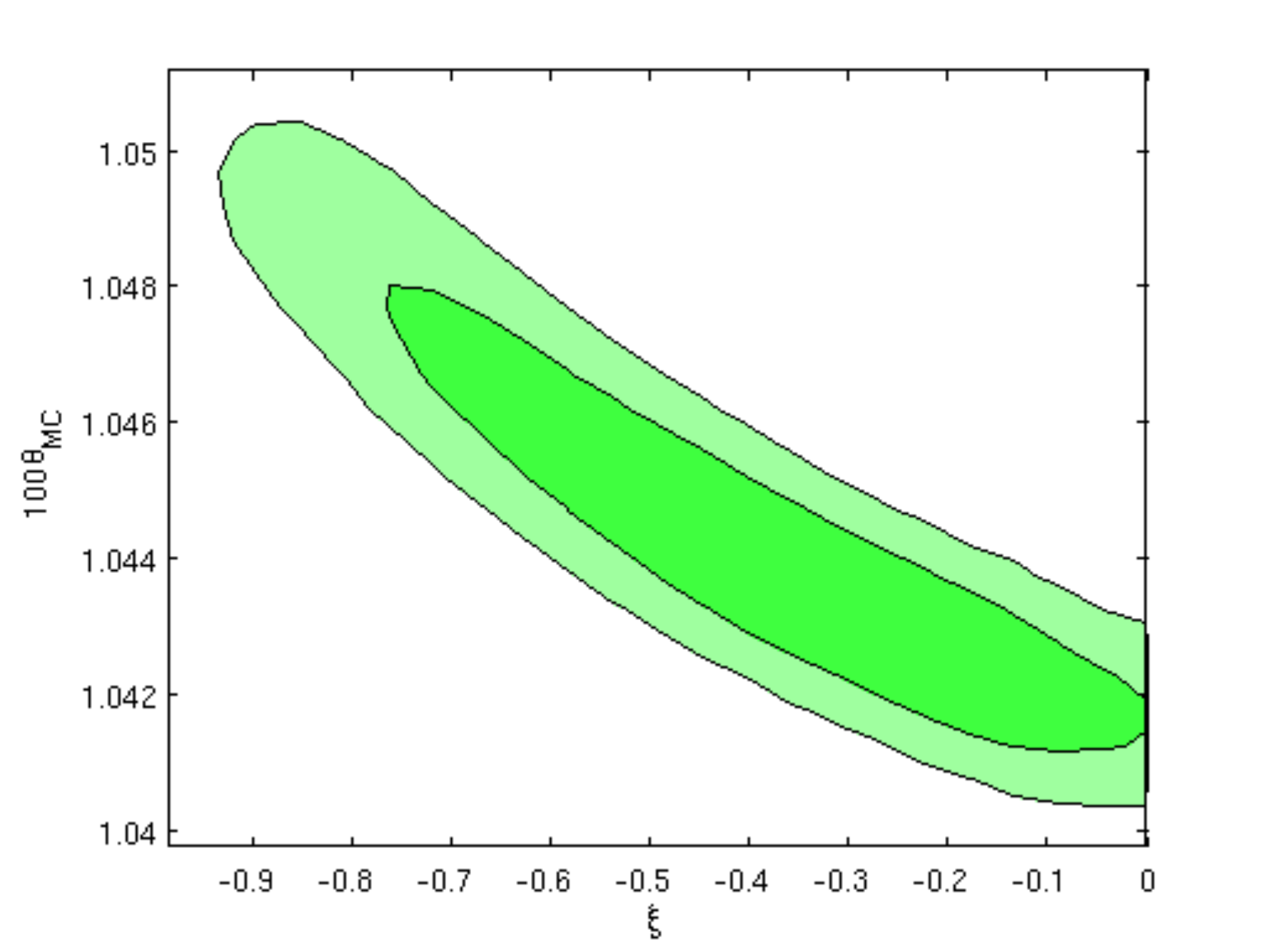}
\includegraphics[width=4.0cm]{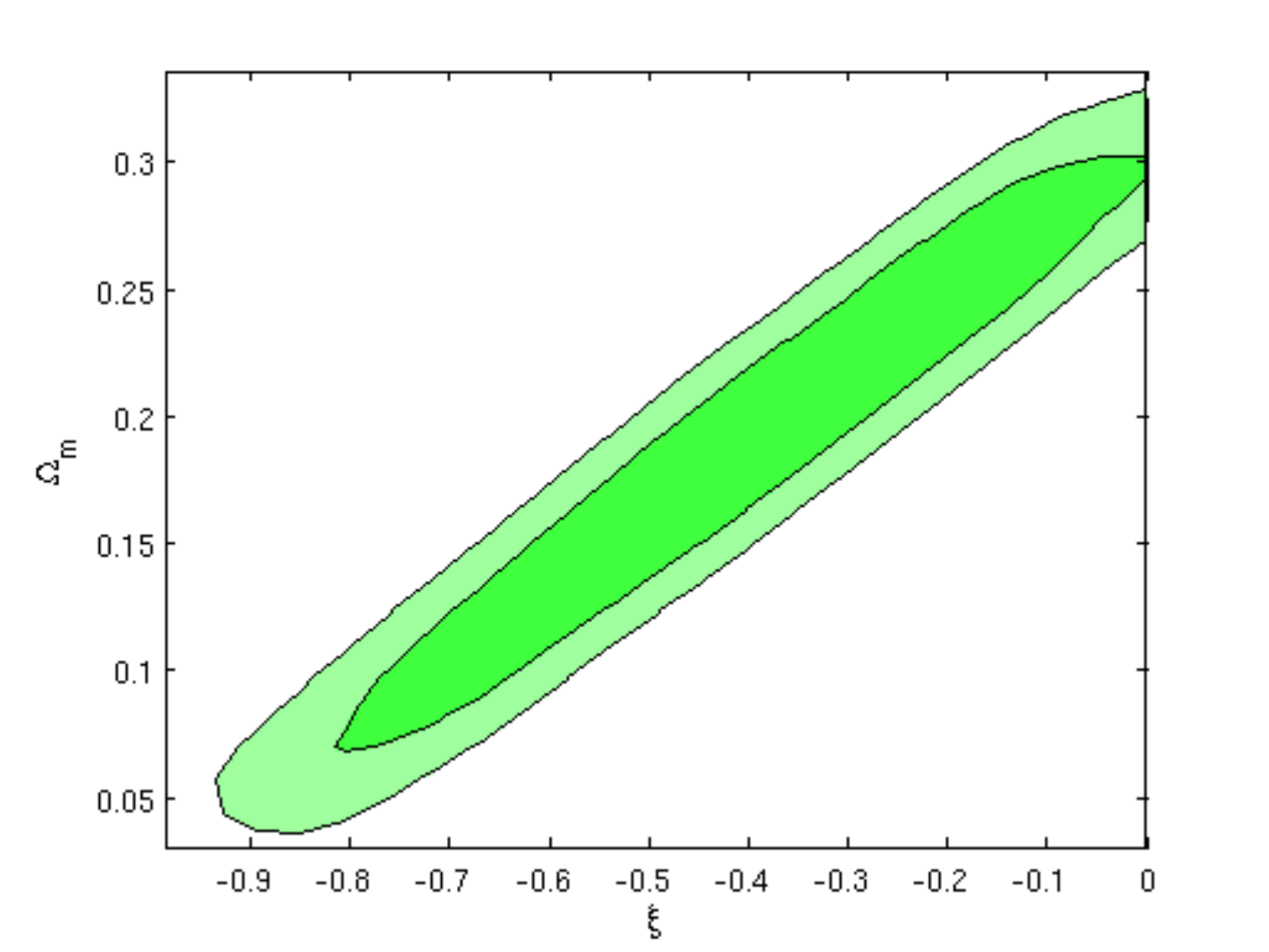}
\includegraphics[width=4.0cm]{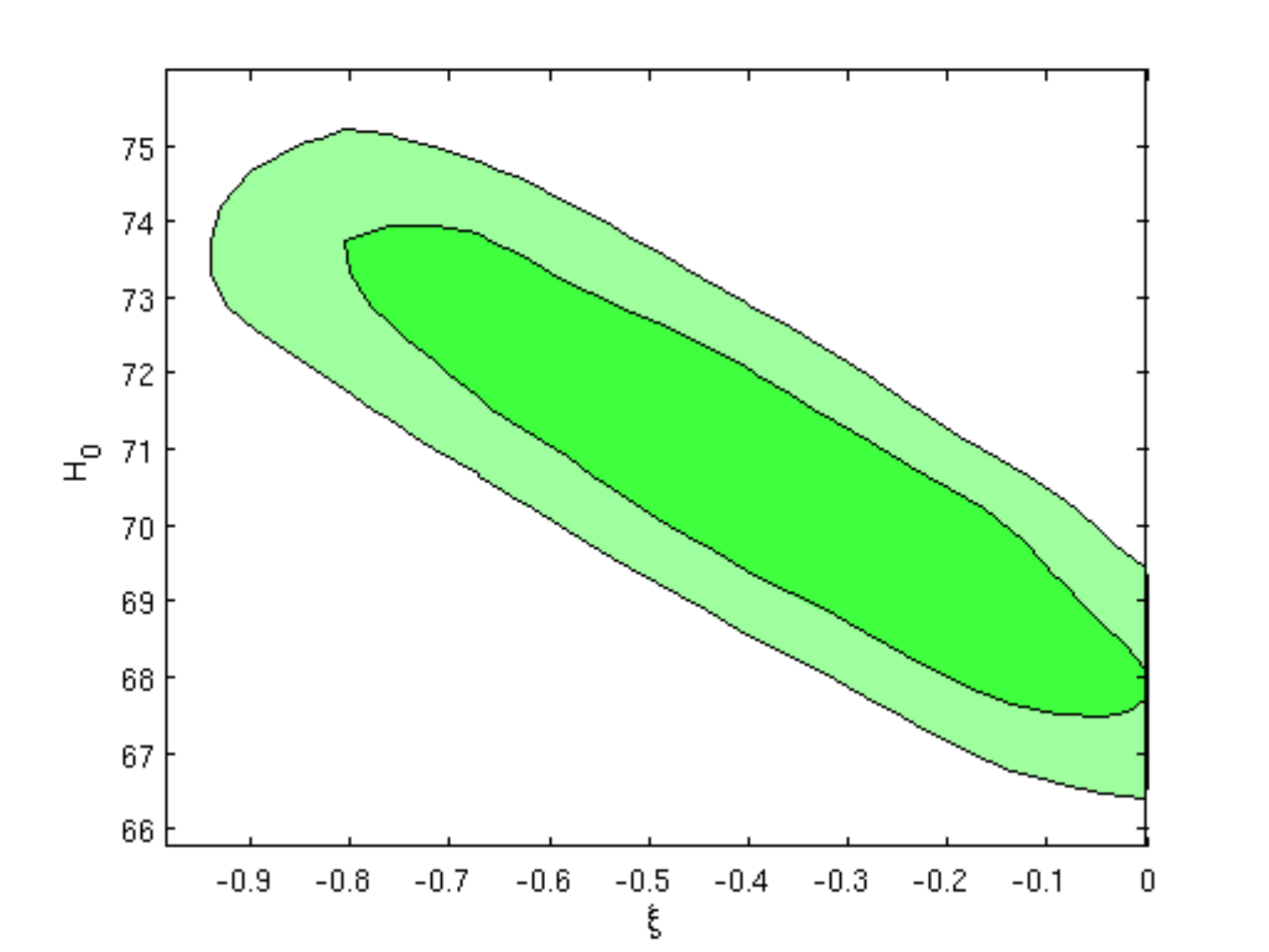}
\caption{2-D posterior distributions  in presence of the HST prior for the same parameters shown in Fig.\ref{fig.2Dcontour}.}
\label{fig.2DcontourBAO}
\end{figure*}

The constraints on the parameters and the best fit values are reported in Tab.~\ref{Tab1}
while we plot the 1-D posteriors for the parameters in Fig.~\ref{fig.posterior_wfix}
and the 2-D posterior for the main parameter degeneracies in Fig.~\ref{fig.2Dcontour},~\ref{fig.2DcontourHST},
\ref{fig.2DcontourBAO}.
Note that the coupling is only between cold
dark matter and dark energy, therefore the bounds on the baryon density remain
unchanged when $\xi$ is allowed to freely vary.

The presence of a dark coupling is perfectly compatible with the
PLANCK data set. Coupled cosmologies provide even a slightly better best
fit $\chi^2$ than the $\Lambda$CDM model, see Tab.~\ref{Tab1}.  
Looking at the marginalized value of $\xi$ in Tab.~\ref{Tab1} it seems to exist also 
a preference for $\xi<0$ but this is mainly due to the large number of coupled dark models 
compatible with the data. The PLANCK data set alone does not exclude $\xi =0$ \ref{fig.posterior_wfix}.

As expected (see e.g~\cite{Honorez:2010rr}), there exists a strong degeneracy between 
$\xi$ and the cold dark matter density $\Omega_c h^2$, see Fig.~\ref{fig.2Dcontour}. 
Negative values of the coupling $\xi$ implies a larger matter density in the
past. Since the PLANCK data are sensitive to the amount of cold dark matter density 
at recombination, $\xi<0$ leads to a lower value of the cold dark matter density today.
The degeneracy is nearly perfect and for large negative values of $\xi$ the PLANCK data set
is compatible with even negligible values of $\Omega_c h^2$.

This degeneracy affects the geometrical parameters values and, in particular, 
the Hubble constant value, which in this case assume values significantly larger than in 
the standard $\Lambda$CDM case. The introduction of a coupling $\xi$ changes the constraint 
of $H_0=67.3\pm1.2$ km/s/Mpc (at $68 \%$ c.l.), obtained in the $\Lambda$CDM case, to $H_0=72.1_{-2.3}^{+3.2}$ km/s/Mpc (at $68 \%$ c.l.).

It is therefore not surprising that, when the HST prior is included, 
the PLANCK+HST data combination suggests a value for the coupling different from zero.
The best fit values and the $68\%$ c.l. constraints we have obtained in this case are reported 
in the third column of table~\ref{Tab1} while 2-D posteriors are shown in Fig.~\ref{fig.2DcontourHST}.
The combined PLANCK+HST constraint excludes a zero value of the coupling parameter $\xi$ at more than 2 sigma ($\xi <-0.22$ at $95\%$ c.l.),
while the value of the Hubble constant is $H_0=73.3_{-1.6}^{+2.6}$ km/s/Mpc at $68 \%$ c.l..

If we add instead the BAO low-redshift measurements to the Planck data, we can observe that a zero coupling it
is admitted but not favoured and that the tension between the Hubble constant measurements is still
alleviated, as in the Planck alone case. The results of this run are shown in the fourth column of 
table~\ref{Tab1} and in Fig.~\ref{fig.2DcontourBAO}.

We have verified that none of the foreground parameters is sensitive to the coupling component, 
confirming that our analysis is not biased by the foreground marginalization.

We have also explored the effect of a freely varying dark energy equation of state $w$ in our results. 
If $w$ is added in the MonteCarlo analyses, the parameters constraints are comparable and the same degeneracies appear, even if in this case the
constraint on $H_0$ is weaker when we consider PLANCK alone. This is due to the fact that CMB measurements alone can weakly constrain $w$ in general,
also in the $\Lambda$CDM case, and can not break the degeneracy between the coupling parameter $\xi$ and $w$. 

\begin{figure*}[htb!]
\centering
\includegraphics[width=8cm]{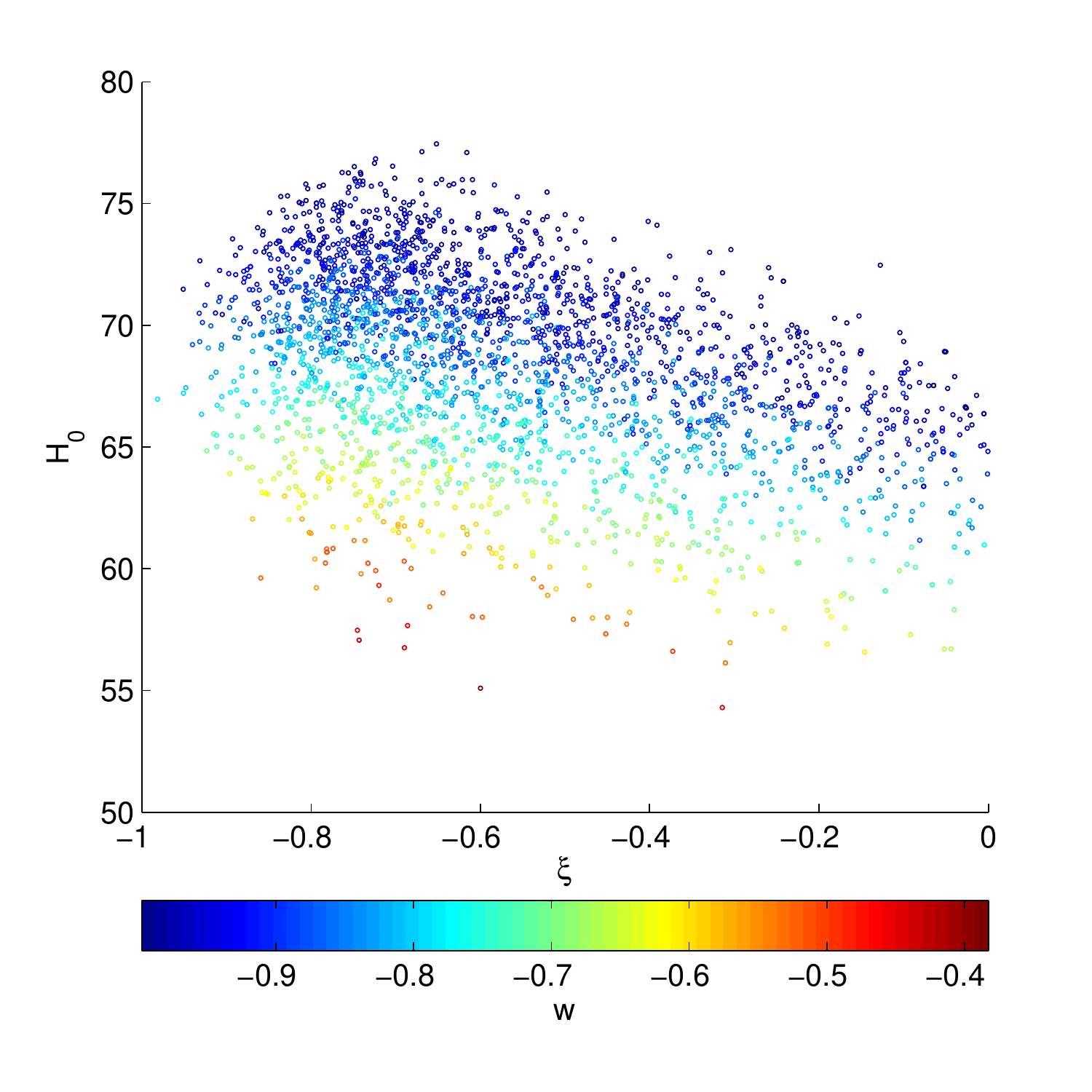}
\caption{3D degeneracy between the coupling parameter $\xi$, $H_0$ and
  $w$ from a run for the PLANCK data set with varying $w$. Note that the values of $H_0$ obtained for reasonable values of
  $w\sim -1$ are in a much better agreement with low redshift measurements of the Hubble constant, if compared to the values of $H_0$ obtained in the pure $\Lambda$CDM PLANCK case.}
\label{fig.H0w}
\end{figure*}

The degeneracy between the coupling 
parameter $\xi$, $H_0$ and $w$ when we consider a varying $w$ for PLANCK data set is shown in Figure~\ref{fig.H0w}. Note that the values of $H_0$ 
obtained for reasonable values of $w\sim -1$ are in a much better agreement with low redshift measurements of the Hubble constant, if compared to 
the values of $H_0$ obtained in the $\Lambda$CDM PLANCK case.

\begin{figure*}[htb!]
\centering
\includegraphics[width=18cm]{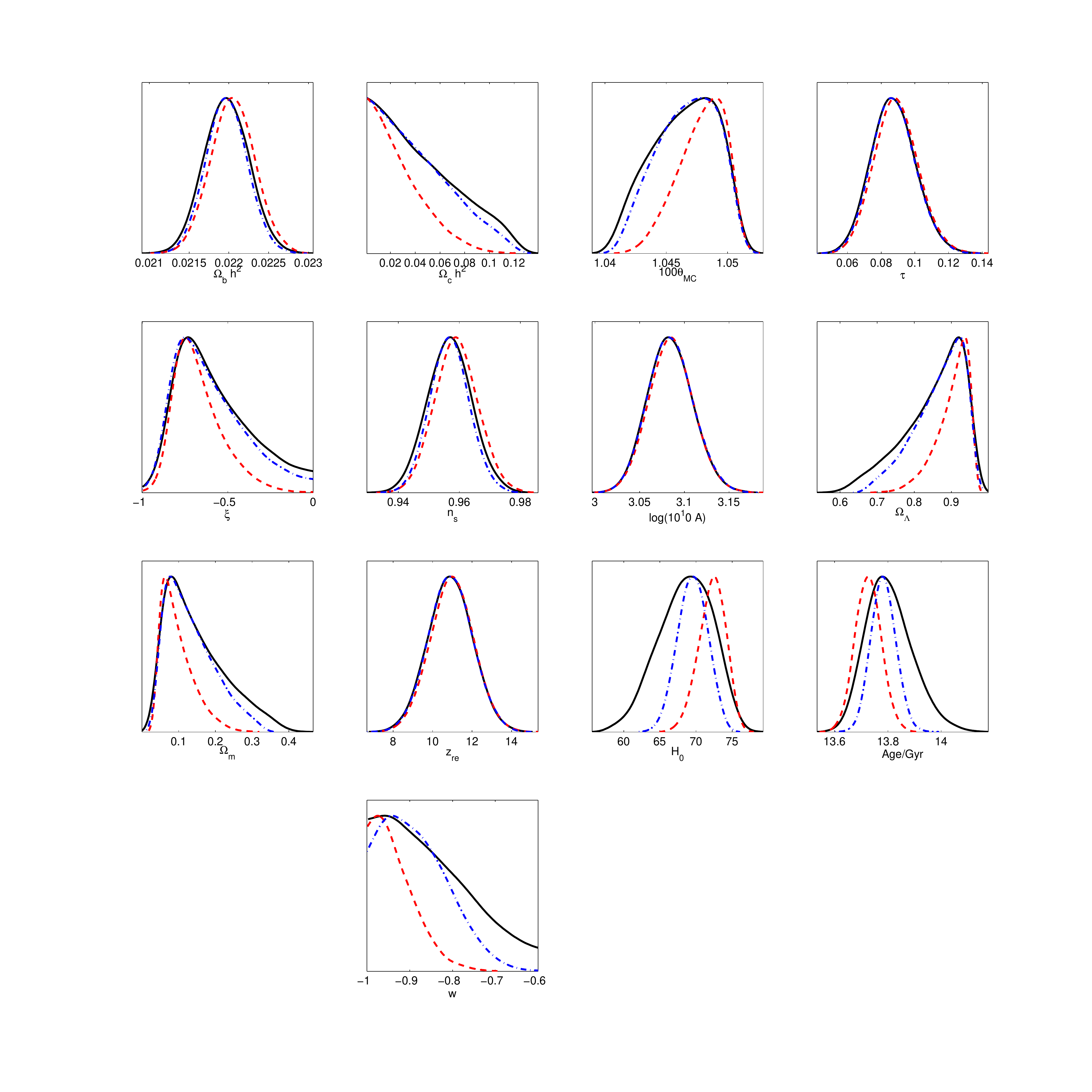}
\caption{Posterior distributions for the cosmological parameters presented in Tab.\ref{Tab2} from PLANCK data set alone (solid black line), PLANCK plus HST prior (red dashed line) and PLANCK plus BAO measurements (blue dot-dashed line).}
\label{fig.posterior_wvar}
\end{figure*}

\begin{figure*}[htb!]
\includegraphics[width=4.0cm]{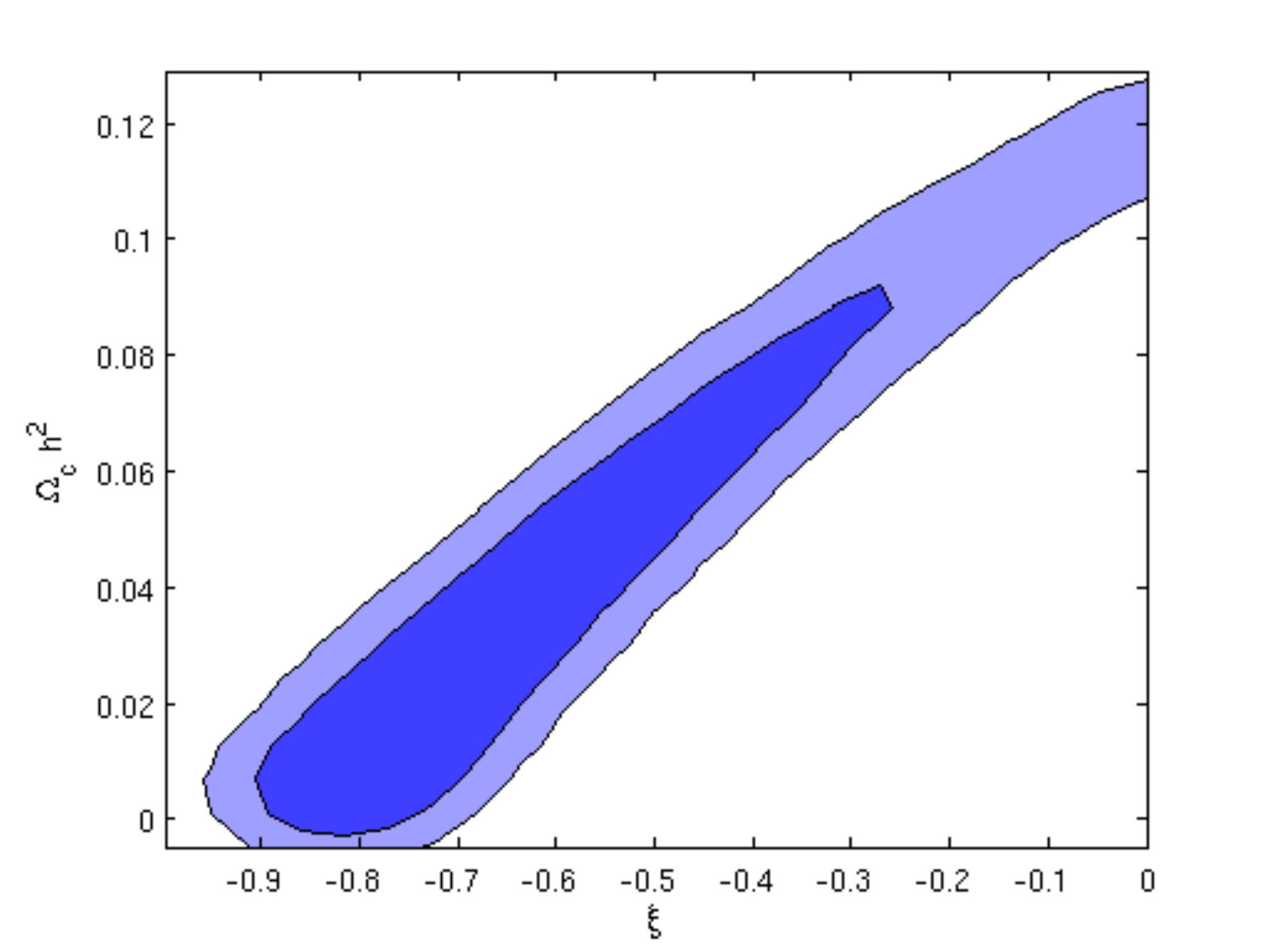}
\includegraphics[width=4.0cm]{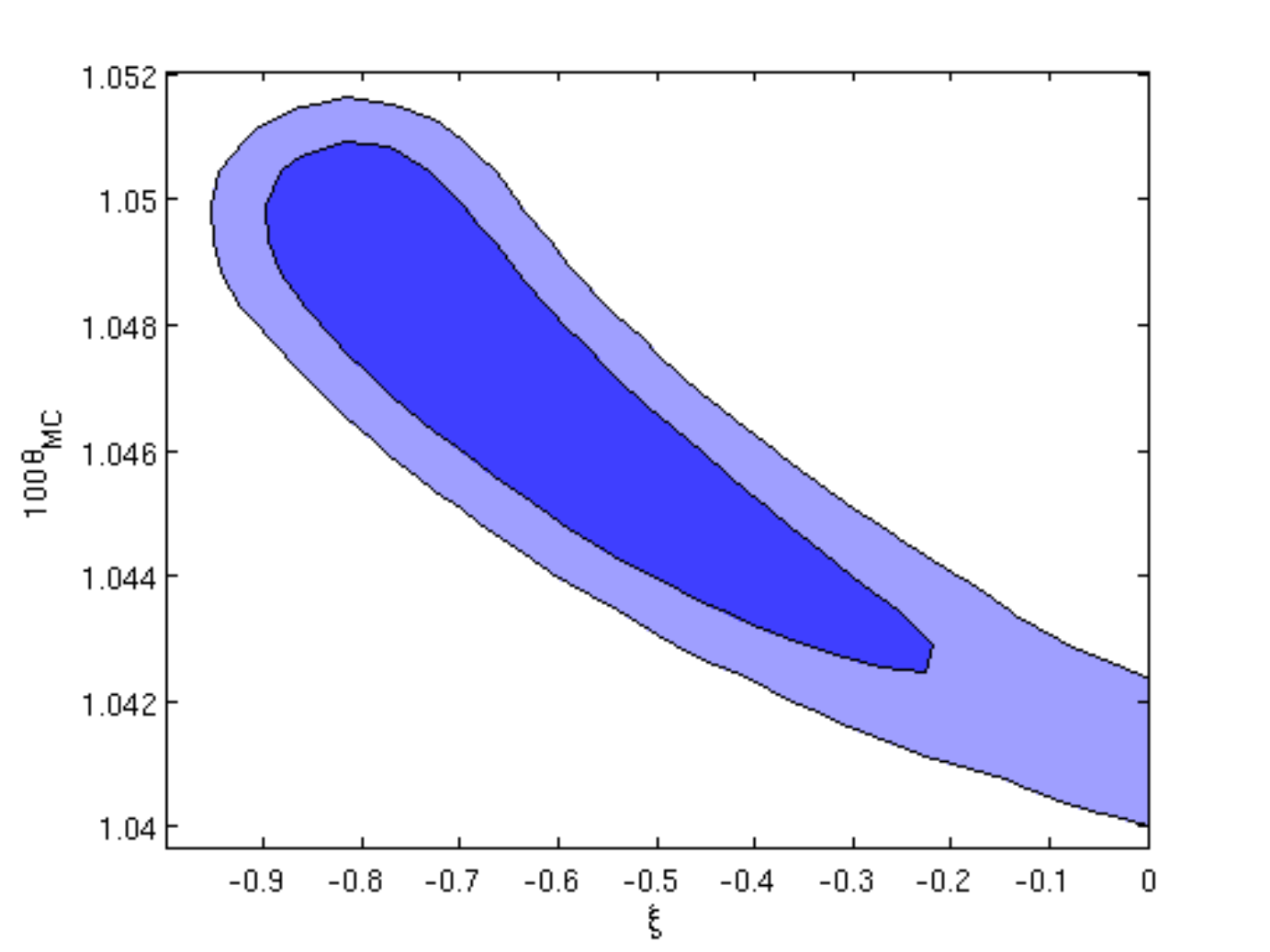}
\includegraphics[width=4.0cm]{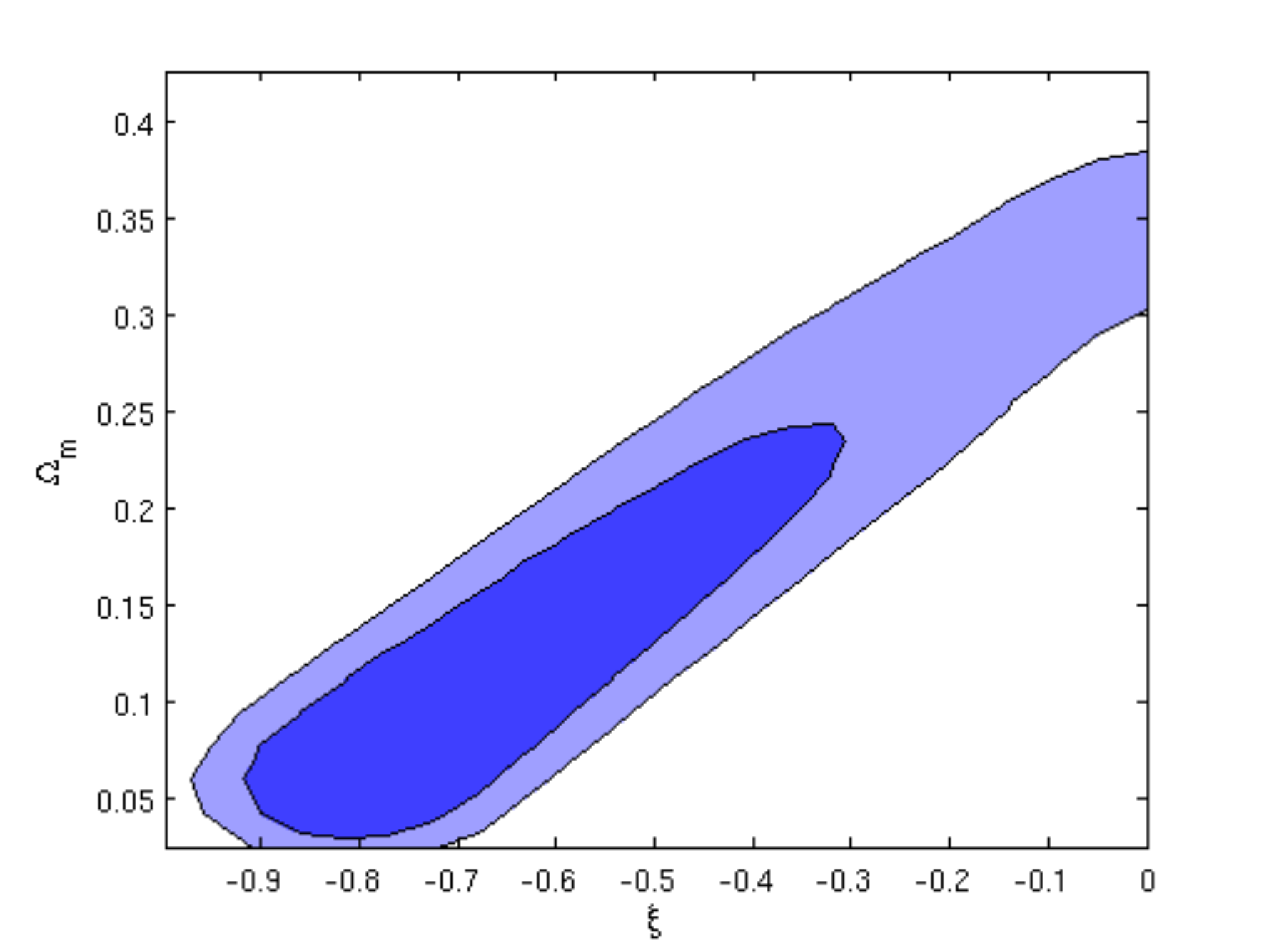}
\includegraphics[width=4.0cm]{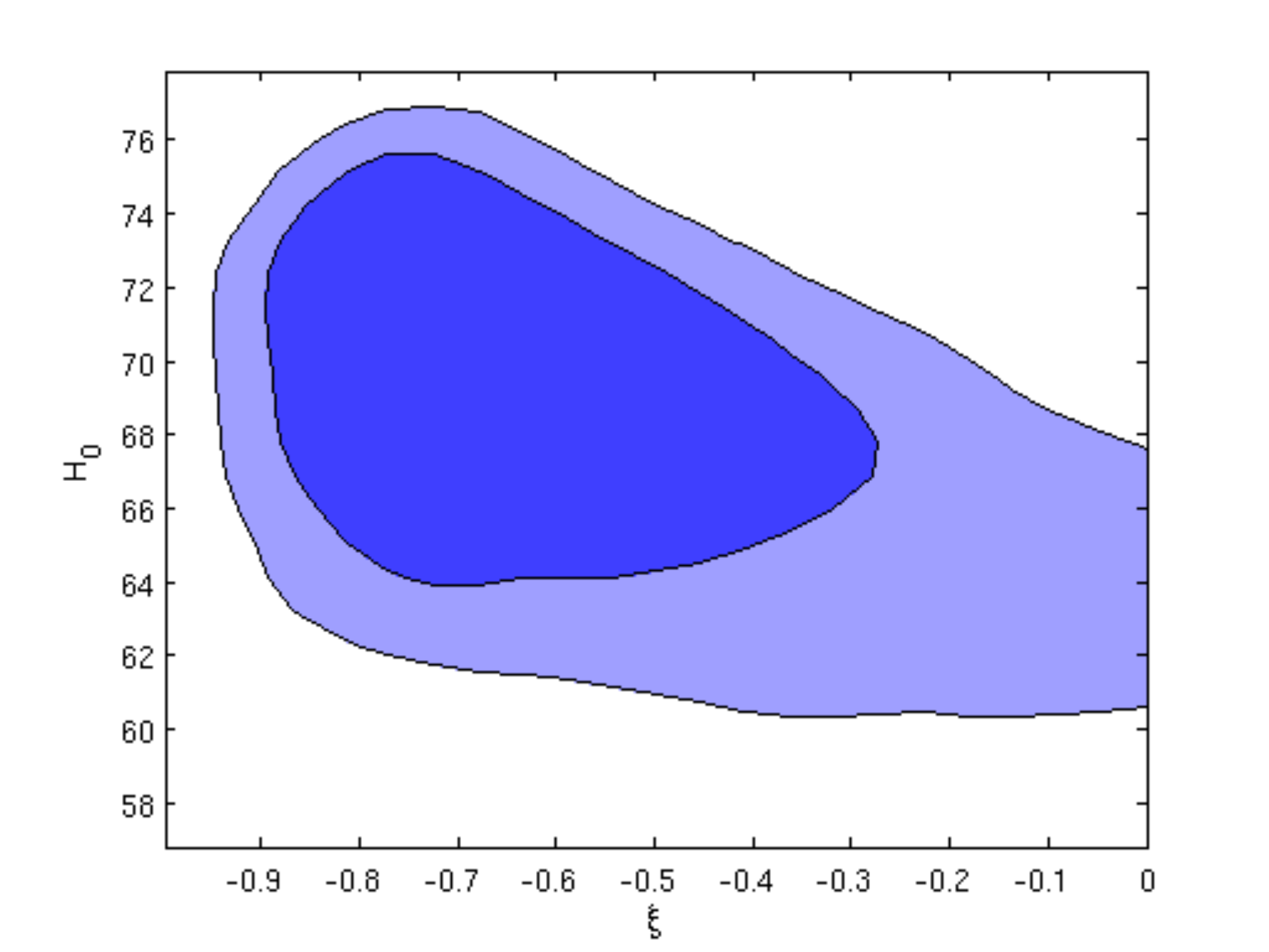}
\caption{2-D posterior distributions of parameters most degenerate with the coupling $\xi$. A strong correlation is evident with the cold dark matter density parameter. A larger absolute value of the coupling $\xi$ implies a decrement of the dark cold matter and a consequent decrease of the dark matter density. Since the assumption of a flat universe, it also implies a larger dark energy amount that brings to an increment of $H_0$ and consequently an increase of $\theta$.}
\label{fig.2Dcontour_wvar}
\end{figure*}

\begin{figure*}[htb!]
\includegraphics[width=4.0cm]{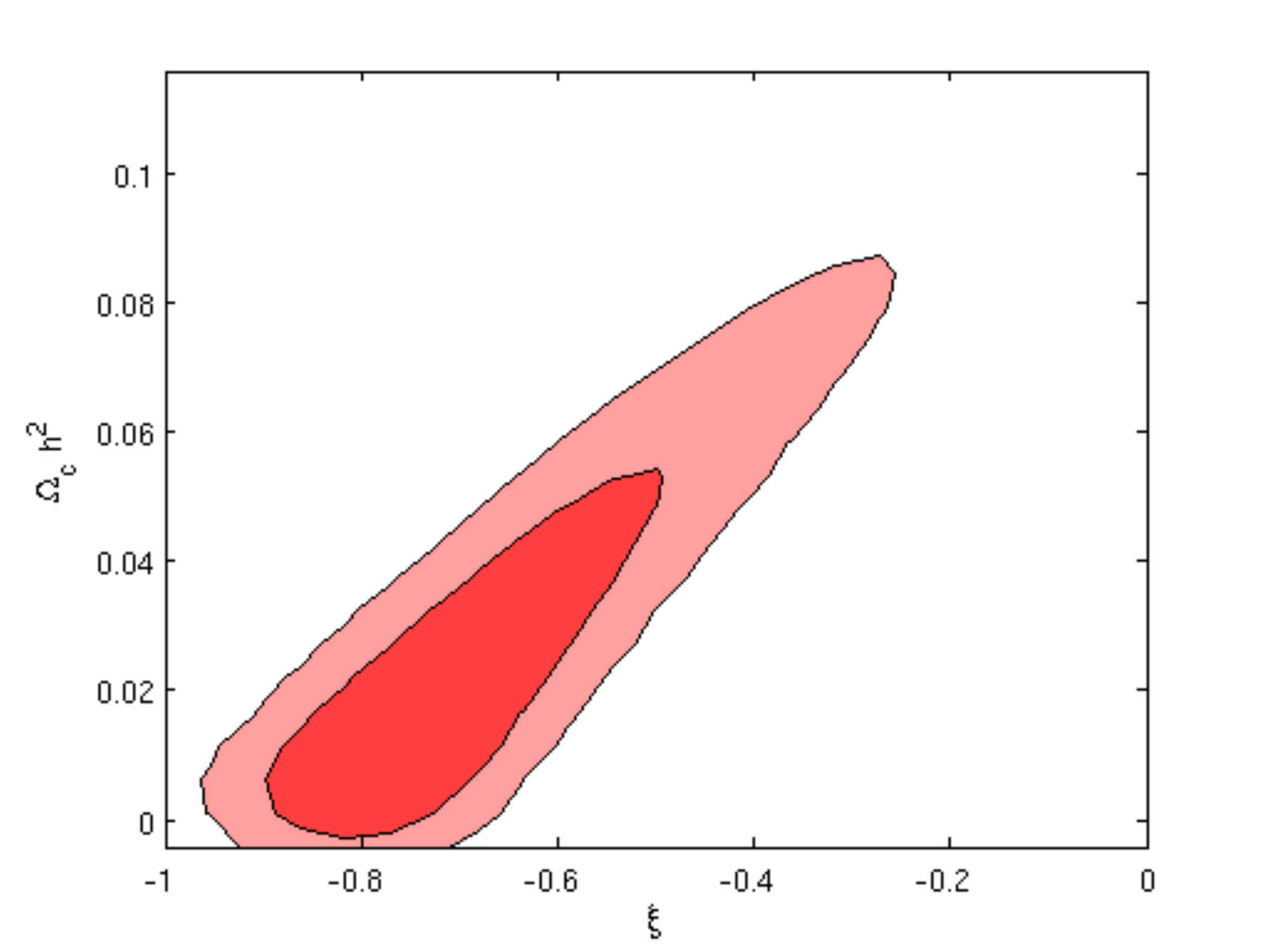}
\includegraphics[width=4.0cm]{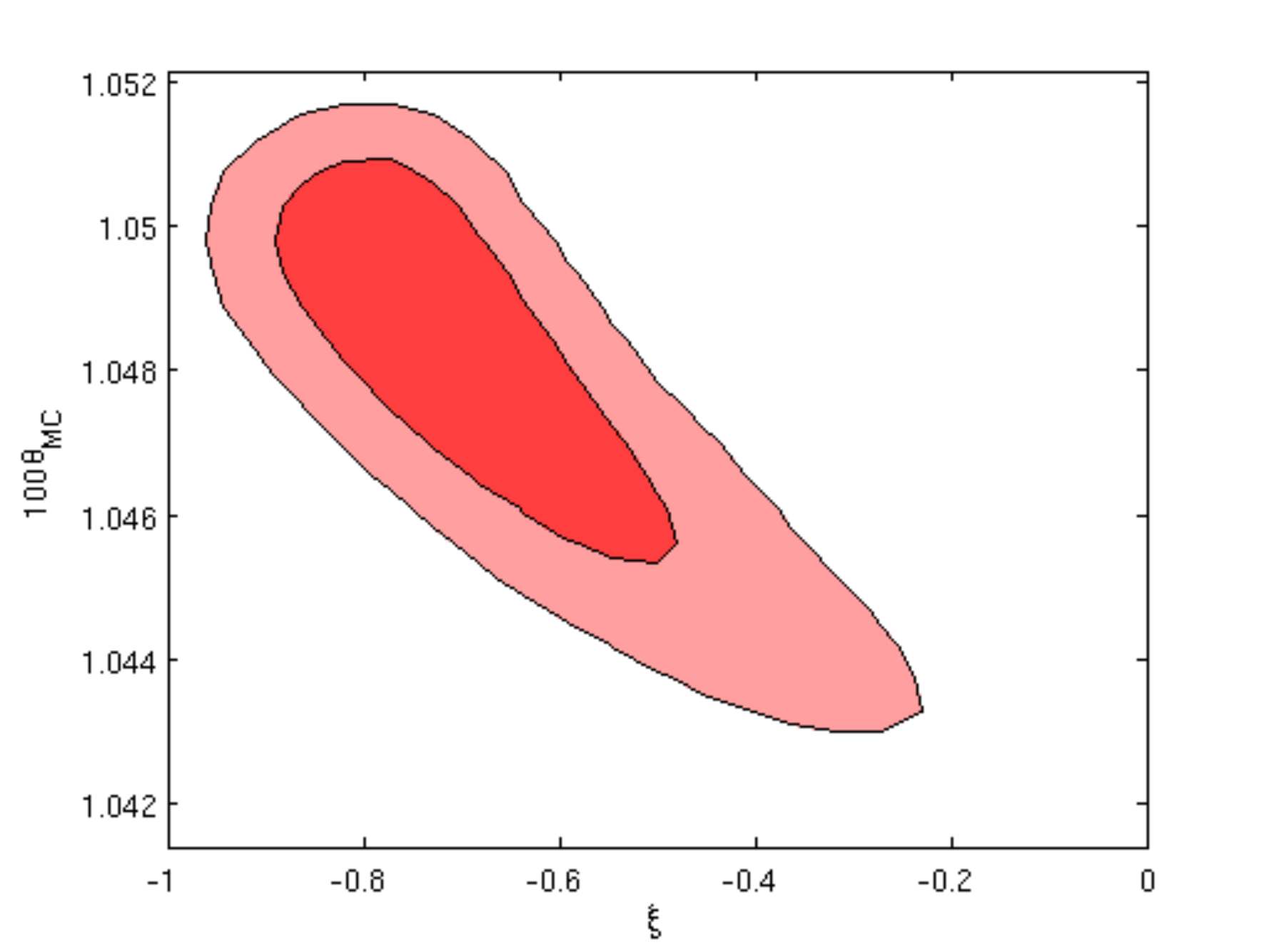}
\includegraphics[width=4.0cm]{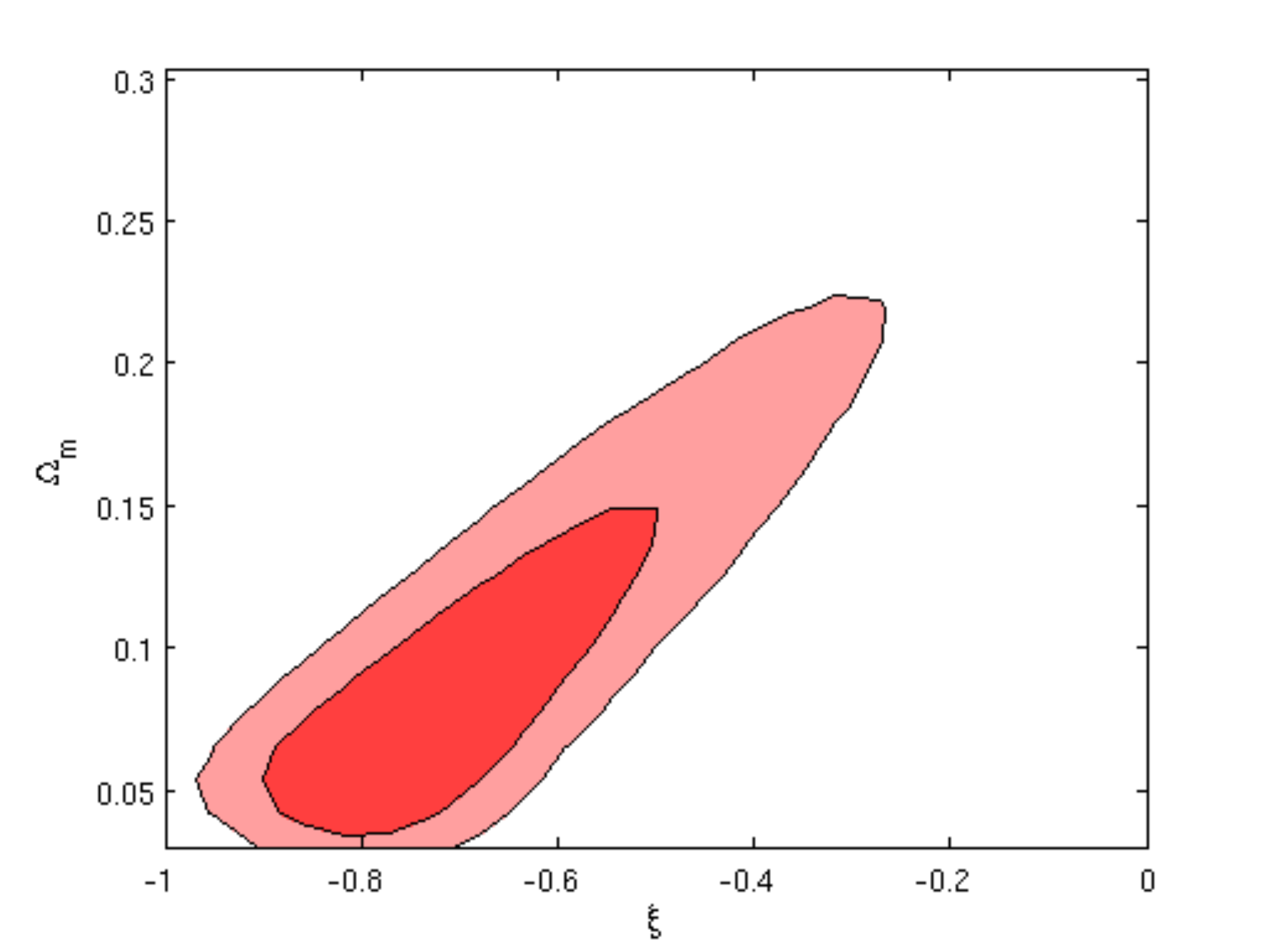}
\includegraphics[width=4.0cm]{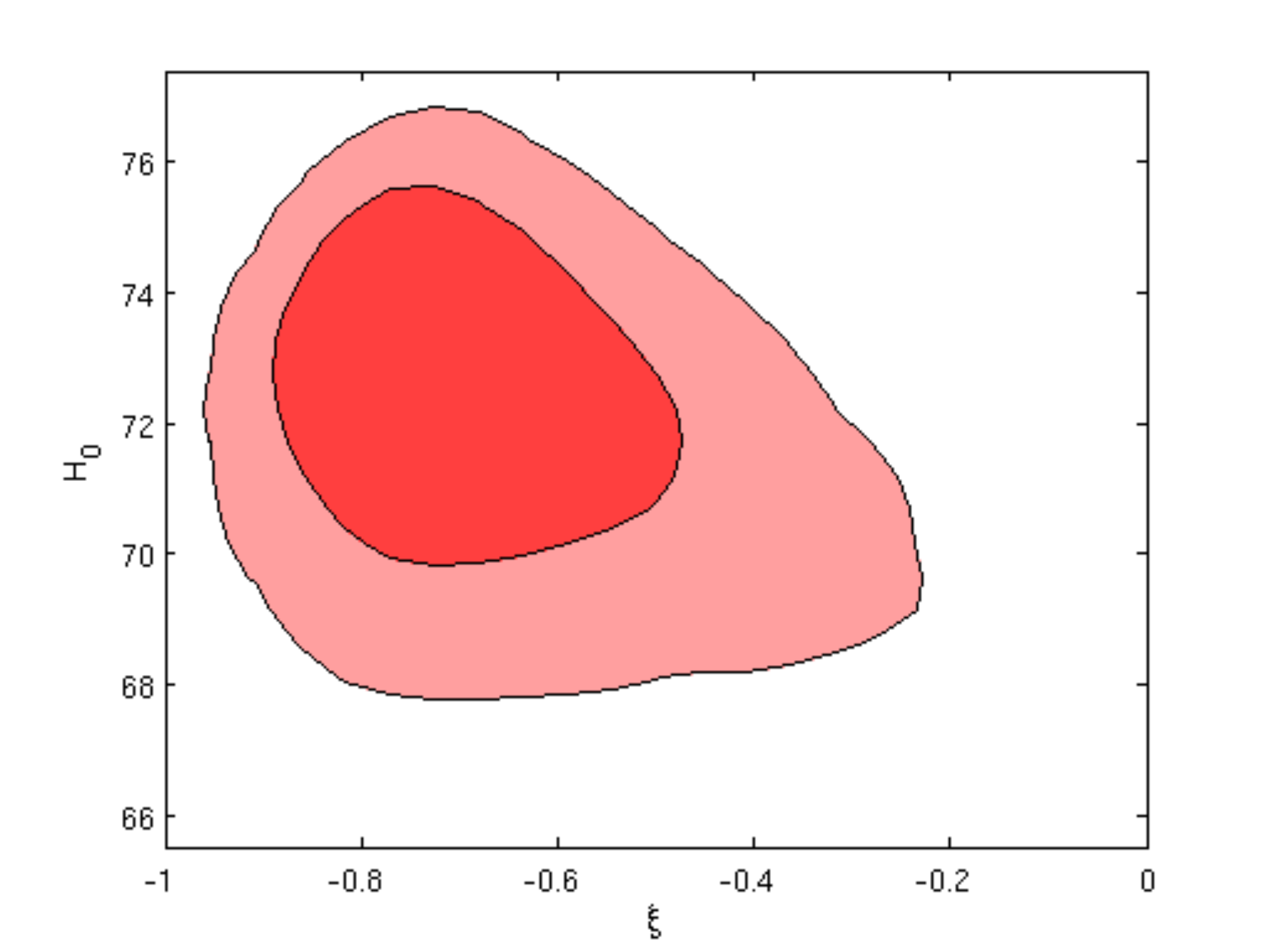}
\caption{2-D posterior distributions  in presence of the HST prior for the same parameters shown in Fig.\ref{fig.2Dcontour}.}
\label{fig.2DcontourHST_wvar}
\end{figure*}

\begin{figure*}[htb!]
\includegraphics[width=4.0cm]{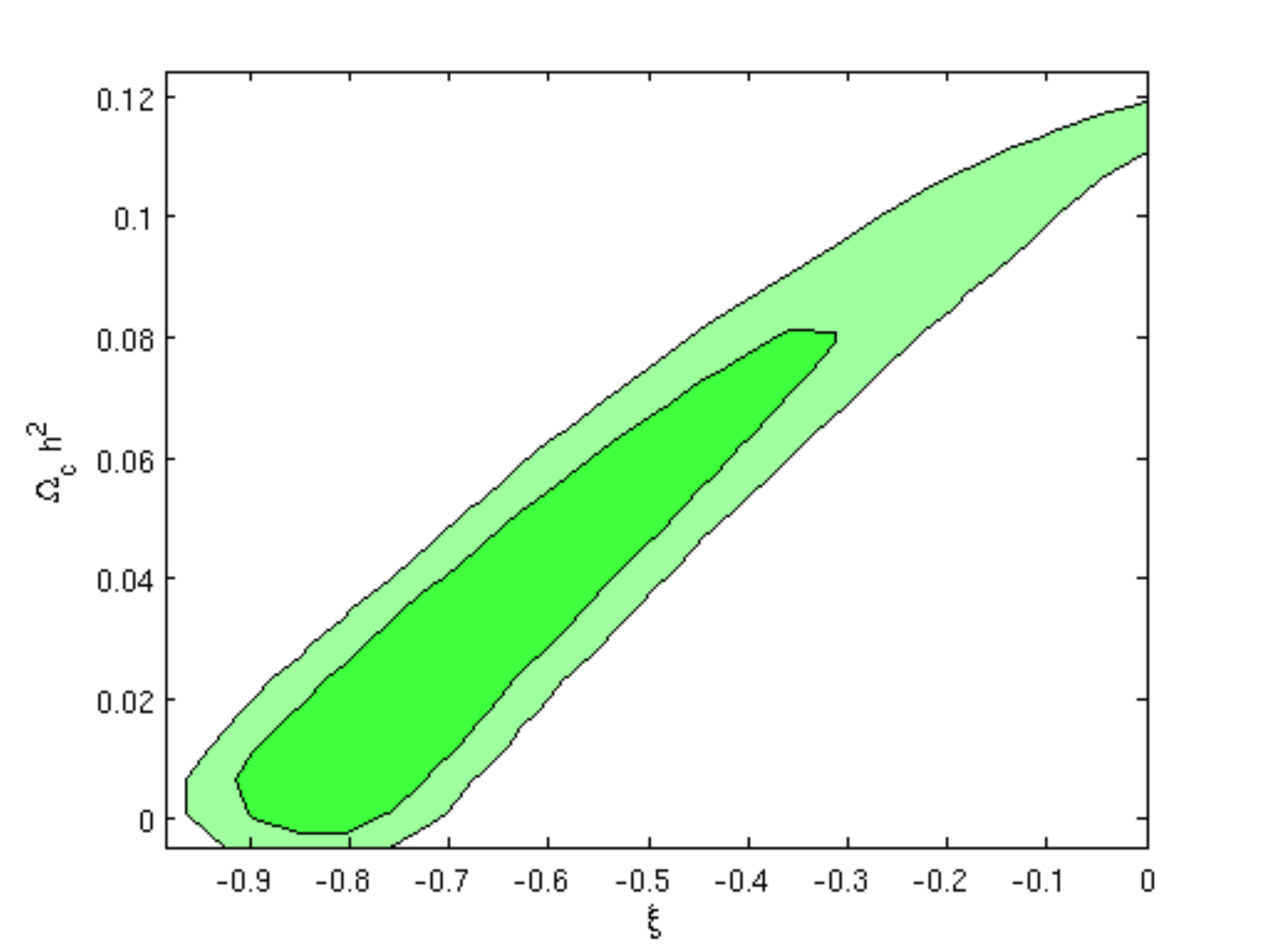}
\includegraphics[width=4.0cm]{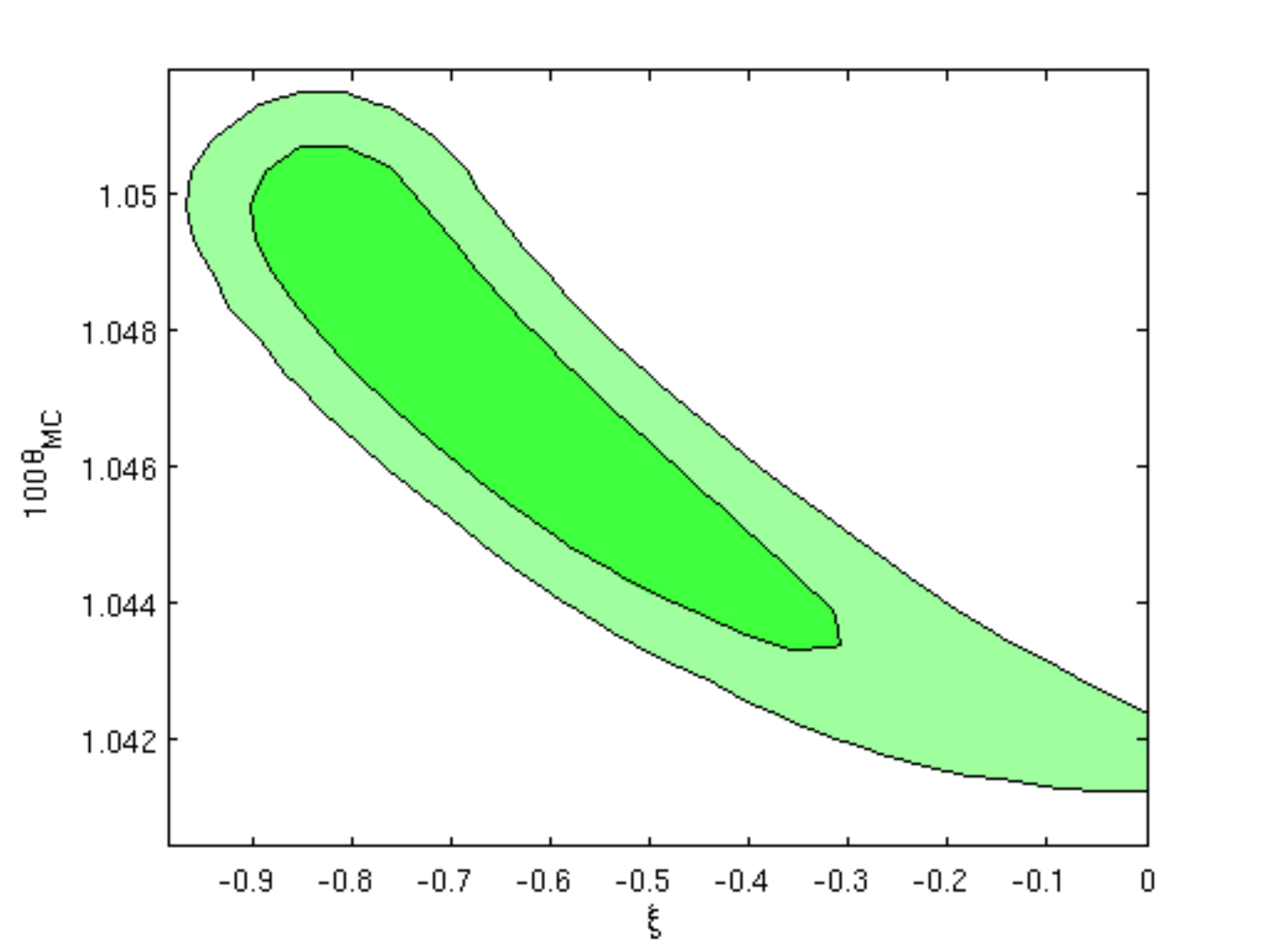}
\includegraphics[width=4.0cm]{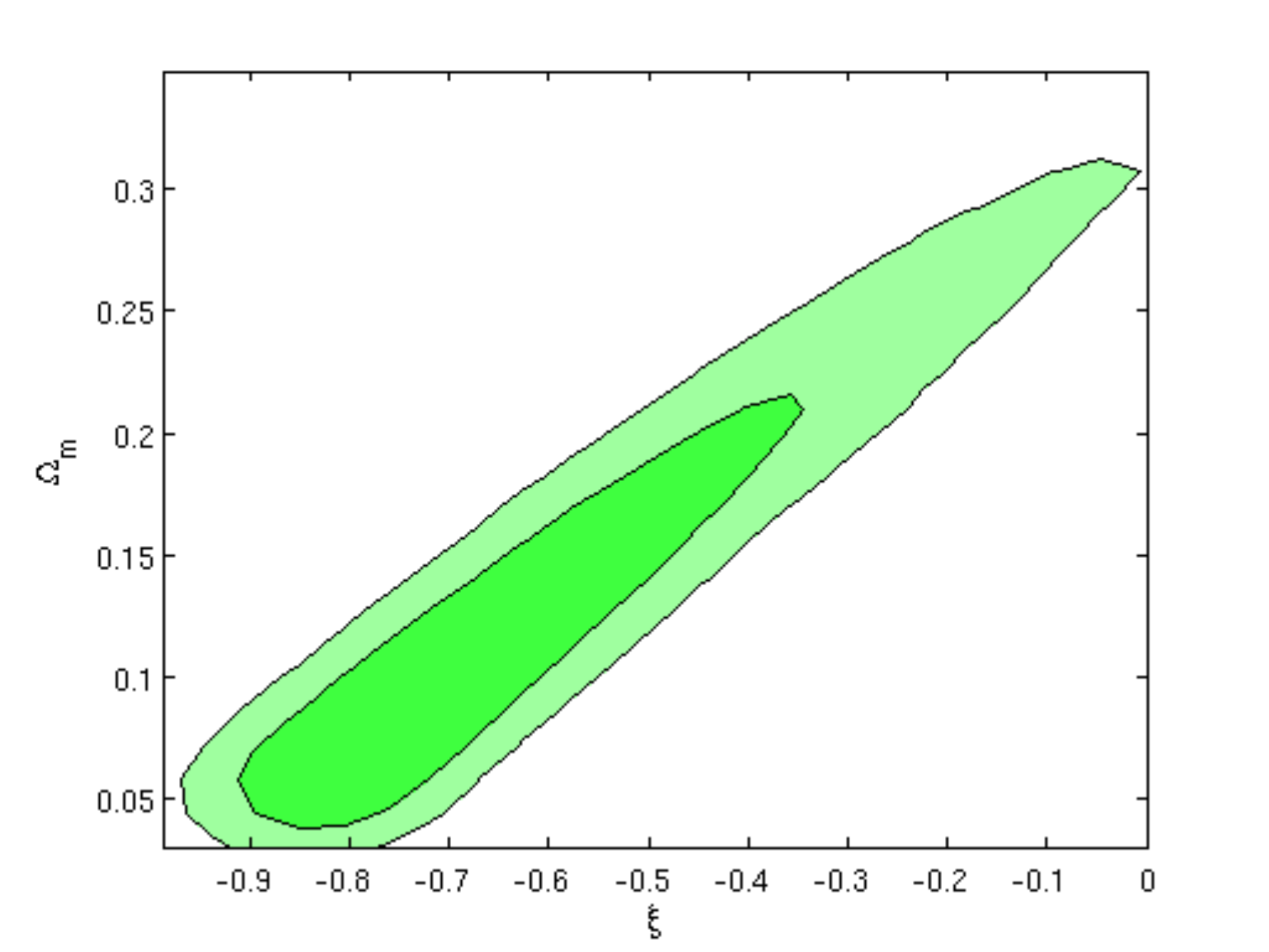}
\includegraphics[width=4.0cm]{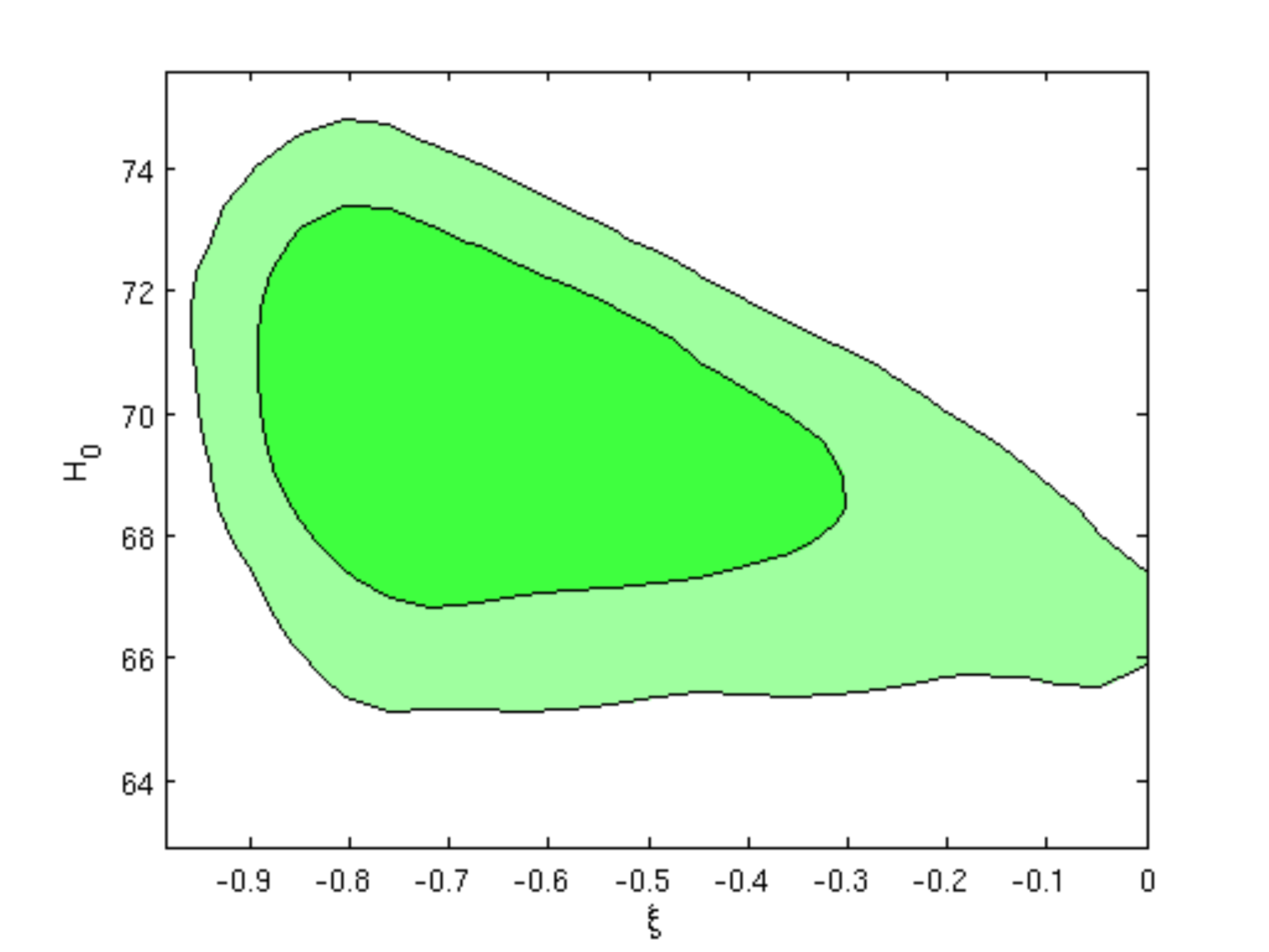}
\caption{2-D posterior distributions  in presence of the HST prior for the same parameters shown in Fig.\ref{fig.2Dcontour}.}
\label{fig.2DcontourBAO_wvar}
\end{figure*}

The best fit values and the $68\%$ c.l. constraints from the three dataset combinations when $w$ is marginalized over are shown in the table~\ref{Tab3}.
While the 1D and 2D posterior distributions are presented in Fig.\ref{fig.posterior_wvar} and~\ref{fig.2Dcontour_wvar},~\ref{fig.2DcontourHST_wvar},~\ref{fig.2DcontourBAO_wvar}.

\begin{table*}[htb!]
\begin{center}
\begin{tabular}{|l||c c|c c||c c||c c|}
\hline
\hline
& \multicolumn{2}{c}{{\bf PLANCK} + {\bf $\Lambda$CDM}} & \multicolumn{2}{|c|}{{\bf PLANCK}+ {\bf $\xi$}} & \multicolumn{2}{|c|}{{\bf PLANCK}+ {\bf HST}+ {\bf $\xi$}} & \multicolumn{2}{|c|}{{\bf PLANCK}+ {\bf BAO}+ {\bf $\xi$}}\\ \hline
Parameters & Best fit & $68\%$ limit &  Best fit & $68\%$ limit  & Best fit & $68\%$ limit & Best fit & $68\%$ limit\\ \hline
$\Omega_b h^2$ &0.02203 & 0.02205 $\pm$ 0.00028 &  0.02204 & 0.02200 $\pm$ 0.00027 & 0.02201  & 0.02203 $\pm$ 0.00027 & 0.02195 &0.02192 $\pm$ 0.00025 \\
$\Omega_c h^2$ &0.1204 & 0.1199 $\pm$ 0.0027 & 0.0664 &$< 0.074$ & 0.0450 & $< 0.056$ & 0.096 & 0.$0.069^{+0.040}_{-0.022}$\\
$100\theta$ & 1.04119 & 1.04131 $\pm$ 0.00063 & 1.0445 & 1.0456 $\pm$ 0.0026 & 1.0461 & 1.0466 $\pm$ 0.0021 & 1.0427 &1.0445 $\pm$ 0.0021\\
$\tau$ &0.093 & 0.089$^{+ 0.012}_{-0.014}$ & 0.0867 &0.087$^{+0.012}_{-0.014}$ & 0.080 & 0.088 $^{+0.017}_{-0.014}$& 0.090& $0.085^{+0.012}_{-0.013}$ \\
$n_s$ &0.9619 & 0.9603 $\pm$ 0.0073 & 0.9543 &0.9580 $\pm$ 0.0071 & 0.9586 & 0.9589 $\pm$ 0.0070& 0.9596&0.9556 $\pm$ 0.0060\\
$\log(10^{10} A_s)$ & 3.098 & 3.089$^{+ 0.024}_{-0.027}$ & 3.090 & 3.083$^{+0.023}_{-0.025}$ & 3.070 & 3.084$^{+0.024}_{-0.027}$&3.090&$3.082^{+0.023}_{-0.026}$\\
$\xi$ & ------& ------& -0.46 &$-0.49^{+0.19}_{-0.31}$ & -0.58 & -0.58$^{+0.090}_{-0.22}$&-0.22&$-0.42^{+0.29}_{-0.21}$\\
\hline
$\Omega_{\rm m}$ & 0.318 & 0.315$^{+0.016}_{-0.018}$ & 0.177 &0.155$^{+0.050}_{-0.11}$ & 0.127 & 0.122$^{+0.033}_{-0.070}$&0.246& $0.187^{+0.085}_{-0.063}$\\
$\Omega_{\rm \Lambda}$ & 0.682  & $0.685^{+ 0.018}_{-0.016}$ & 0.823 & 0.845$^{+0.11}_{-0.050}$ & 0.873 & 0.878$^{+0.070}_{-0.033}$&0.754& $0.813^{+0.063}_{-0,085}$\\
$z_{re}$ & 11.4 & $11.1 \pm 1.1 $ & 10.9 &10.9  $\pm$ 1.1 & 10.2 & 10.9 $\pm$ 1.1&11.2&10.7 $\pm$ 1.1\\
$H_0 [\mathrm{km}/\mathrm{s}/\mathrm{Mpc}]$ & 67.0 & 67.3 $\pm$ 1.2 & 71.0 & 72.1$^{+3.2}_{-2.3}$ & 73.0 & 73.3$^{+2.0}_{-1.6}$&69.3&$70.8^{+1.9}_{-2.1}$\\
$Age/Gyr$ & 13.824 & 13.817 $\pm$ 0.048 & 13.747 &$13.733^{+0.062}_{-0.065}$ & 13.720 & 13.711$^{+0.051}_{-0.046}$&13.785&13.765$\pm$ 0.044\\
\hline
\hline
$\chi^2_{\rm min}/2$ &\multicolumn{2}{c}{4902.95} & \multicolumn{2}{|c|}{4902.45} & \multicolumn{2}{|c|}{4902.52} & \multicolumn{2}{|c|}{4902.71} \\
\hline
\hline
\end{tabular}
\caption{Best fit values and 68$\%$ c.l. constraints for the $\Lambda$CDM model (first column) and the dark coupled model described in the text from PLANCK data set (second column). We can see that the effect of the coupling, encoded by the parameter $\xi$, is to increase $H_0$ and to decrease $\Omega_c h^2$ respect to the standard model values. If we consider the combined constraint on the dark coupled model from PLANCK and HST (third column) we see similar bounds but now a negative coupling is significantly favoured. If we instead combine PLANCK and BAO measurements (fourth column) we observe that a lower value of $\Omega_c h^2$ is favoured, compared to coupled model in the PLANCK alone case, but the $H_0$ value is still in agreement with the Hubble Space Telescope measurement.}
\label{Tab1}
\end{center}
\end{table*}

\begin{table*}[htb!]
\begin{center}
\begin{tabular}{|l||c c|c c||c c||c c|}
\hline
\hline
& \multicolumn{2}{c}{{\bf PLANCK} + {\bf w$\Lambda$CDM}} & \multicolumn{2}{|c|}{{\bf PLANCK}+ {\bf $\xi$}} & \multicolumn{2}{|c|}{{\bf PLANCK}+ {\bf HST}+ {\bf $\xi$}} & \multicolumn{2}{|c|}{{\bf PLANCK}+ {\bf BAO}+ {\bf $\xi$}}\\ \hline
Parameters & Best fit & $68\%$ limit &  Best fit & $68\%$ limit  & Best fit & $68\%$ limit & Best fit & $68\%$ limit\\ \hline
$\Omega_b h^2$ &0.02220 & 0.02206 $\pm$ 0.00028 &  0.02207 & 0.02197 $\pm$ 0.00028 & 0.02207623  & 0.02205 $\pm$ 0.00027& 0.02192 &0.02197 $\pm$ 0.00025\\
$\Omega_c h^2$ &0.1200 & 0.1199 $\pm$ 0.0026 & 0.0607 &$< 0.058$ & 0.013 & $< 0.034$ & 0.012 & $<0.052$\\
$100\theta$ & 1.04125 & 1.04132 $\pm$ 0.00063 & 1.0445 & 1.0465 $\pm$ 0.0026 & 1.0492 & 1.0479$\pm$ 0.0019 & 1.0487 &1.0467 $\pm$ 0.0023\\
$\tau$ &0.090 & 0.089$^{+ 0.012}_{-0.014}$ & 0.093 &0.087$^{+0.012}_{-0.014}$ & 0.094& 0.089 $^{+0.0087}_{-0.013}$& 0.085& $0.088^{+0.012}_{-0.013}$ \\
$n_s$ &0.9598 & 0.9602 $\pm$ 0.0071 & 0.9614 &0.9568 $\pm$ 0.0071 & 0.9582 &  0.9590 $\pm$  0.0070& 0.9563&0.9570 $\pm$ 0.0064\\
$\log(10^{10} A_s)$ & 3.093 & 3.089$^{+ 0.024}_{-0.027}$ & 3.095 & 3.084$^{+0.024}_{-0.025}$ & 3.098 & 3.086 $\pm$ 0.024&3.086&$3.085^{+0.023}_{-0.026}$\\
$\xi$ & ------& ------& -0.48 &$-0.58^{+0.13}_{-0.28}$ & -0.76 & -0.67$^{+0.086}_{-0.17}$&-0.82&$-0.61^{+0.12}_{-0.25}$\\
w & -1.92 & $-1.49^{+0.25}_{-0.43}$&  -0.870& $-0.855^{+0.041}_{-0.14}$ & -0.952 &$-0.932^{+0.35}_{-0.067}$ &  -0.887&$-0.883^{+0.033}_{-0.12}$\\
\hline
$\Omega_{\rm m}$ & 0.147 & 0.218$^{+0.023}_{-0.079}$ & 0.179 &0.148$^{+0.034}_{-0.11}$ & 0.065 & $0.098^{+0.017}_{-0.055}$& 0.131& $0.187^{+0.032}_{-0.083}$\\
$\Omega_{\rm \Lambda}$ & 0.853  & $0.682^{+ 0.079}_{-0.023}$ & 0.821 & 0.852$^{+0.11}_{-0.034}$ & 0.935 & 0.902$^{+0.055}_{-0.017}$&0.869& $0.813^{+0.083}_{-0.032}$\\
$z_{re}$ & 11.1 & $11.0 \pm 1.1 $ & 11.4 &10.9  $\pm$ 1.1 & 11.5 & 11.0 $\pm$ 1.1&10.8&10.9 $\pm$ 1.1\\
$H_0 [\mathrm{km}/\mathrm{s}/\mathrm{Mpc}]$ & 98.6 & 83$^{+17}_{-5}$ & 68.2 & 68.5$^{+4.4}_{-3.3}$ & 73.7 & 72.2$^{+2.1}_{-1.6}$&70.8&69.6 $\pm$ 2.0\\
$Age/Gyr$ & 13.449 & 13.817 $\pm$ 0.048 & 13.789 &$13.609^{+0.080}_{-0.18}$ & 13.696 & 13.726 $\pm$ 0.048&13.770&13.783 $\pm$ 0.046\\
\hline
\hline
$\chi^2_{\rm min}/2$ &\multicolumn{2}{c}{4903.14} & \multicolumn{2}{|c|}{4902.02} & \multicolumn{2}{|c|}{4901.90} & \multicolumn{2}{|c|}{4902.59} \\
\hline
\hline
\end{tabular}
\caption{Best fit values and 68$\%$ c.l. constraints for the $\Lambda$CDM model (first column) and the dark coupled model described in the text from PLANCK data set alone (second column), PLANCK+HST (second column) and PLANCK+BAO (third column) when we marginalize over the dark energy equation of state parameter $w$. It is important to note that the prior on $w$ in the coupled model ($w>-1$) is not the same of the $\Lambda$CDM case. As we can expect the constraints are weaker in this case compared to Tab.~\ref{Tab1}, since both $w$ and $xi$ depends on $\rho_{de}$, but the same trends are evident. Also note that the constraints on the background evolution parameters, as $H_0$ or $w$, from CMB measurements alone are very large also in the $\Lambda$CDM model.}
\label{Tab3}
\end{center}
\end{table*}

\section {Conclusions} \label {sec:concl}

In this paper we have presented novel cosmological constraints on a
dark matter-dark energy interaction from the new CMB measurements provided by the
Planck experiment. We have found that a dark coupling interaction is
compatible with Planck data and that the coupling parameter $\xi$ is weakly
constrained by Planck measurements  $\xi=-0.49^{+0.19}_{-0.31}$. 
However, the inclusion of a dark coupling opens up new degeneracies
and affects strongly the constraints on the remaining cosmological parameters.
The model with the dark interaction gives a lower matter density
$\Omega_{\rm m}=0.155^{+0.050}_{-0.11}$ 
and a larger Hubble parameter $H_0=72.1^{+3.2}_{-2.3}\mathrm{km}/\mathrm{s}/\mathrm{Mpc}$. 
Since the value of the Hubble constant is compatible with the HST value, we
have combined the Planck and HST data sets, finding that, in this
case, a non-zero value of the dark coupling is suggested by the data, with $-0.90< \xi <-0.22$ at $95\%$ c.l..
The analysis presented here points out that an interaction in the dark sector is not
only allowed by current CMB data but 
can even resolve the tension between the Planck and the HST measurements of the Hubble parameter. 
The results we have found are in agreement with the results obtained in 
former analyses for similar models using previous
cosmological data \cite{Honorez:2010rr,Pettorino:2012ts,Valiviita:2009nu,Clemson:2011an}.

\subsection*{Acknowledgements}
It is a pleasure to thank Eleonora Di Valentino, Martina Gerbino and Najla Said  for useful discussions and help.
\appendix
\section{Adiabatic initial conditions} \label{appendix}
In ref.~\cite{Gavela:2010tm}, the evolution equations for the coupled
dark matter dark energy model considered here were studied in a gauge
invariant way. It was demonstrated that, assuming adiabatic initial
conditions for all the standard cosmological fluids (photon,
baryons,...), the coupled dark energy fluid also obeys adiabatic
initial conditions, see also Ref.~\cite{Doran:2003xq} in the uncoupled
case and Ref.~\cite{Majerotto:2009np} for another coupled example. At leading
order in $x=k\tau$, in the synchronous gauge, the initial conditions read:
\begin{eqnarray}
  \delta_{de}^{in} (x)&=&(1+w-2\xi)\frac{(1+w+\xi/3)}{12w^2-2w-3w \xi+7
    \xi-14}\, 
  \left(\frac{-2\delta_\gamma^{in}(x)}{1+w_\gamma}\right)\,, \cr
  v_{de}^{in}&=&\frac{x(1+w+\xi/3)}{12w^2-2w-3w \xi+7
    \xi-14}\,
  \left(\frac{-2\delta_\gamma^{in}(x)}{1+w_\gamma}\right)\,, \nonumber
\end{eqnarray}
where $\delta_\gamma^{in}(x)$ are the initial conditions for the
photon density perturbations and $w_\gamma=1/3$ is the equation of
state of the photon.  The latter reduce to the adiabatic initial
conditions for dark energy perturbations in the synchronous gauge
obtained in Ref.~\cite{Ballesteros:2010ks} in the uncoupled case.

\label{lastpage}
\end{document}